\documentclass{article}
\usepackage[utf8]{inputenc}
\usepackage[left=3cm,right=3cm,top=3cm,bottom=3cm]{geometry}
\usepackage{pdfpages}
\usepackage{hyperref}
\usepackage{listings}
\usepackage{xcolor}
\usepackage{setspace} 
\usepackage{tocloft}
\usepackage{amsmath}
\usepackage{chngcntr }
\usepackage[toc,page]{appendix}
\usepackage[T1]{fontenc}
\usepackage[nottoc]{tocbibind}
\usepackage[compact]{titlesec}
\usepackage[utf8]{inputenc}
\usepackage[sectionbib]{natbib}
\usepackage{chapterbib}    
\setcitestyle{aysep={,}}

\usepackage{multirow}
\usepackage{setspace}
\usepackage{array}

\newcolumntype{L}[1]{>{\raggedright\let\newline\\\arraybackslash\hspace{0pt}}m{#1}}
\newcolumntype{C}[1]{>{\centering\let\newline\\\arraybackslash\hspace{0pt}}m{#1}}
\newcolumntype{R}[1]{>{\raggedleft\let\newline\\\arraybackslash\hspace{0pt}}m{#1}}

\usepackage{subfig}
\usepackage{bm}
\usepackage{multirow}
\usepackage{tikz}
\usetikzlibrary{shapes,arrows}
\usepackage{blkarray}
\usepackage[affil-it]{authblk}

\title{DRHotNet Arxiv}

\begin{document}

\title{DRHotNet: An R package for detecting differential risk hotspots on a linear network}

\author[1,2*]{Álvaro Briz-Redón}
\affil[1]{Department of Statistics and Operations Research, University of València, Spain}

\author[2]{Francisco Martínez-Ruiz}
\affil[2]{Statistics Office, City Council of València, Spain}

\author[1]{Francisco Montes}

\maketitle

\abstract{One of the most common applications of spatial data analysis is detecting zones, at a certain investigation level, where a point-referenced event under study is especially concentrated. The detection of this kind of zones, which are usually referred to as hotspots, is essential in certain fields such as criminology, epidemiology or traffic safety. Traditionally, hotspot detection procedures have been developed over areal units of analysis. Although working at this spatial scale can be suitable enough for many research or practical purposes, detecting hotspots at a more accurate level (for instance, at the road segment level) may be more convenient sometimes. Furthermore, it is typical that hotspot detection procedures are entirely focused on the determination of zones where an event is (overall) highly concentrated. It is less common, by far, that such procedures prioritize the location of zones where a specific type of event is overrepresented in relation to the other types observed, which have been denoted as differential risk hotspots. The R package DRHotNet provides several functionalities to facilitate the detection of differential risk hotspots along a linear network. In this paper, DRHotNet is depicted and its usage in the R console is shown through a detailed analysis of a crime dataset. 
}

\section{Introduction}

Hotspot detection basically consists in finding zones of a space where certain event is highly concentrated. There exists a wide variety of methods in literature that allow researchers to identify hotspots at a certain level of accuracy or spatial aggregation. Some of them have been massively used in the last decades, including certain local indicators of spatial association such as LISA \citep{anselin1995local} or the Getis-Ord statistic \citep{getis1992analysis}, and the spatial scan statistic \citep{kulldorff1997spatial}. The first two of these methods are implemented in the R package \textsf{spdep} \citep{AppliedSpatialDataAnalysisWithR}, whereas the scan statistic is implemented in \textsf{DCluster} \citep{gomez2005detecting} (although there are other R packages that also provide an implementation of these methods). Furthermore, many new R packages focused on hotspot detection have been released in the last years, most of them being model-based and oriented to disease mapping studies that are carried out over administrative (areal) units \citep{allevius2018scanstatistics,JSSv090i14,meyer2017spatio}.

However, the analysis of certain types of events requires detecting hotspots at a level of spatial accuracy greater than that provided by administrative or regular areal units. Indeed, many research studies of the fields of criminology \citep{andresen2017trajectories,weisburd2015law} and traffic safety \citep{briz2019spatial,nie2015network,xie2013detecting} that have been published in recent years were entirely carried out on road network structures rather than on administrative units. More specifically, some quantitative criminologists have estimated that around 60\% of the total variability in crime incidence occurs at the street segment level \citep{schnell2017influence,steenbeek2016action}, a fact that shows the essentiality of using road segments instead of areal structures to capture the spatial concentration of certain events more properly. 

Fortunately, road network structures were introduced in the context of spatial statistics some years ago, providing the basis for analyzing events lying on such structures, which are usually referred to as linear networks. Indeed, a planar linear network, $L$, is defined as a finite collection of line segments, $L=\cup_{i=1}^{n} l_{i}$, in which each segment contains the points $l_{i}=[u_{i},v_{i}]=\{tu_{i}+(1-t)v_{i} : t\in [0,1]\}$ \citep{ang2012geometrically,baddeley2015spatial,baddeley2017stationary}. Following graph theory nomenclature, these segments are sometimes referred to as the edges of the linear network, whereas the points that determine the extremes of such segments are known as the vertices of the network. 

Hence, a point process $X$ on $L$ is a finite point process in the plane such that all points of $X$ lie on the network $L$ \citep{ang2012geometrically,baddeley2015spatial,baddeley2017stationary}. Similarly, a collection of events that is observed on $L$ is known as a point pattern, $x$, in $L$. In addition, when every event of a point pattern has one or several attributes, the point pattern is referred to as a marked point pattern. Each of the attributes, which can be in the form of either a numerical or a categorical variable, is known as a mark of the pattern.

The investigation of spatial patterns lying on linear networks is gaining attention in the last years. The design of new and more accurate/efficient kernel density estimators \citep{mcswiggan2017kernel,moradi2018kernel,moradi2019resample,rakshitfast}, the introduction of graph-related intensity measures \citep{eckardt2018point}, the construction of local indicators of spatial association \citep{eckardt2017second}, or the estimation of relative risks \citep{mcswiggan2019estimation} are some topics that have recently started developing for linear networks.

Besides the theoretical advances, it is worth noting that using linear networks for carrying out a spatial or spatio-temporal analysis entails certain technical difficulties. In this regard, the R package \textsf{spatstat} \citep{baddeley2015spatial} provides multiple specific functions that allow R users to carry out statistical analyses on linear networks. Furthermore, efforts are constantly being made to reduce the computational cost of adapting certain classical spatial techniques to the singular case of linear networks \citep{rakshit2019efficient}.

Despite the existing necessity of analyzing point-referenced data coming from certain fields of research at the street level, there are not many software tools fully designed for hotspot detection on road networks. One relevant contribution in this regard is the KDE+ software \citep{bil2016kde+}, but this is not integrated in R. The package \textsf{DRHotNet} is specifically prepared for allowing R users to detect differential risk hotspots (zones where a type of event is overrepresented) along a linear network.

\section{A procedure for detecting differential risk hotspots}

The procedure for differential risk hotspot detection available in \textsf{DRHotNet} was introduced by \cite{briz2019hotspots}, who also show an application of the method considering a traffic accident dataset. Overall, hotspot detection methods can be classified into partition-, hierarchy- and density- based methods \citep{deng2019density}. The one implemented in \textsf{DRHotNet} belongs to the last of these three groups. Furthermore, from a statistical point of view, the method can be considered to be nonparametric, as no statistical model (with parameters) is involved.

Hence, the following subsections provide a description of each of the steps that are carried out by the \textsf{DRHotNet} package. The specification and exemplification of the functions required for each of them is given in following sections.

\subsection{Estimating a relative probability surface}

The first step consists in using kernel density estimation to infer the relative probability of occurrence for a certain type of event along a linear network. Assuming a marked point pattern $\{x_{1},...,x_{n}\}$, where a binary mark $y_{i}$ indicates if $x_{i}$ corresponds, or not, to the type of event of interest, the following expression \citep{kelsall1998spatial} is used to estimate this relative probability (which depends on kernel's bandwidth parameter, $\sigma$):

\begin{equation} 
\label{eq:EqProbability}
p_{\sigma}(x)=\sum_{i=1}^{n} \lambda_{\sigma}(x_{i})y_{i}\biggl/\sum_{i=1}^{n} \lambda_{\sigma}(x_{i})
\end{equation}
where $\lambda_{\sigma}(x)$ is computed according to the network-constrained equal-continuous kernel density estimation provided by \cite{mcswiggan2017kernel}. This version of kernel density is implemented in the function \texttt{density.lpp} of \textsf{spatstat}, which computes kernel density values rapidly by solving the classical heat equation defined on the linear network, as demonstrated by \cite{mcswiggan2017kernel}.

Hence, $p_{\sigma}(x)$ approximates the relative probability of occurrence, at location $x$, of the type of event being represented by $y_{i}=1$. 
In simpler terms, Equation \ref{eq:EqProbability} allows establishing a relative probability surface that varies along the linear network. To this end, given a location, $x$, of the linear network, the events which mainly contribute to the estimation of $\lambda_{\sigma}(x)$ are those situated within a linear radius (following the linear network structure) of $\sigma$ meters from $x$. Thus, increasing the value of $\sigma$ leads to smoother representations of the probability surface, whereas smaller values of the bandwidth parameter produce the opposite effect. On the choice of an optimal bandwidth parameter, the method described by \cite{mcswiggan2019estimation} could be followed, which modifies previous proposals made for planar patterns \citep{kelsall1995kernel,kelsall1995non}. 

Estimating a relative probability surface implies, in practice, estimating a relative probability value at the middle points of the segments forming the road network. In this regard, it is necessary to split the original road network structure into shorter line segments called lixels \citep{xie2008kernel}, which are basically the analogous to pixels for an areal space. This step provides accuracy and homogeneity to the process of hotspot formation. Then, the segments satisfying certain conditions (details are provided in the next section) are selected and joined together forming the differential risk hotspots. 

\subsection{Determining differential risk hotspots}

Once a relative probability surface has been estimated along the linear network, it is time to detect differential risk hotspots. The procedure for their detection relies on two parameters $k$ and $n$. Parameter $k$ is used together with the standard deviation of all the relative probabilities estimated to define a threshold that needs to be exceeded to consider a segment as a candidate to be part of a differential risk hotspot. Then, parameter $n$ is employed to select those segments, from the ones satisfying the threshold condition, that have $n$ or more events at a distance lower than $\sigma$. More precisely, a segment, $i$, of the linear network is considered to be part of a differential risk hotspot if:

$$\hat{p_{i}} > \mathrm{mean}(\{\hat{p_{j}}\}_{j=1}^{S})+ k \cdot \mathrm{sd}(\{\hat{p_{j}}\}_{j=1}^{S})$$

$$\#\{x \in \{x_{1},...,x_{n}\} : d_{L}(x,m_{i})<\sigma \}\geq n$$
where $S$ is the number of segments of the linear network, $\hat{p_{i}}$ is the relative probability of occurrence for the type of event of interest estimated at the middle point of segment $i$, $\#$ denotes the cardinality of a set, $m_{i}$ is the middle point of segment $i$ and $d_{L}(x,m_{i})$ represents the shortest-path distance (distance along the network) between $x$ and $m_{i}$. 

The segments that fulfil the two conditions indicated above are then joined, forming differential risk hotspots. More precisely, two segments satisfying the aforementioned conditions belong to the same differential risk hotspot if they are neighbours. Given two segments of a linear network, a neighbouring relationship exists between them if they share a vertex of the network, which is equivalent to the \textit{queen} criterion used for defining polygon neighbourhoods \citep{lloyd2010spatial}. This relationship between segments is referred to as a first-order one. Similarly, higher-order neighbouring relationships are defined recursively (for instance, for a second-order relationship): $i$ and $j$ are second-order neighbours if one neighbour of $i$ and one neighbour of $j$ are first-order neighbours.

\section{Measuring differential risk hotspots importance and significance}

\subsection{The prediction accuracy index}

Given a collection of hotspots identified in a planar space, \cite{chainey2008utility} defined the prediction accuracy index (PAI) as follows:

\begin{equation} \label{eq:EqPAI}
\mathrm{PAI}=\frac{\mathrm{Hit\;rate}}{\mathrm{Area\;percentage}}=\frac{n/N}{a/A}
\end{equation}
where $n$ is the total number of events that lie on the hotspots, $N$ is the total number of events observed, $a$ is the area formed by all the hotspots together, and $A$ is the total area of the space. Hence, a higher value of PAI is convenient from the perspective of hotspot detection, as it represents a higher concentration of the events in relation to the extension of the physical space that they occupy. This formula can be easily adapted to the case of linear networks using segment lengths in the place of areas. 

In the context of detecting hotspots where the relative probability of one type of event is particularly high, the PAI needs to be modified. A type-specific version of the PAI (denoted by $\mathrm{PAI}_{type}$) was proposed in \cite{briz2019hotspots} (for linear networks):

\begin{equation} 
\label{eq:PAItype}
\mathrm{PAI}_{type}=\frac{n_{type}/N_{type}}{l/L}
\end{equation}
where $n_{type}$ is the number of events of the type of interest that lie on the hotspots, $N_{type}$ is the total number of events of that type that are observed, $l$ is the length of all the hotspots together, and $L$ is the total length of the network. The interpretation of the $\mathrm{PAI}_{type}$ is analogous to the one for the PAI, that is, a higher $\mathrm{PAI}_{type}$ value indicates a higher proportion of the type of interest in relation to certain proportion of road network length. Some researchers have recently explored the possibility of using the number of segments (in total and within the hotspots) instead of segment length \citep{drawve2019research} in the denominator of this formula, but we opted for segment length proportions.

It is also worth noting that the $\mathrm{PAI}_{type}$ can be computed for the whole set of differential risk hotspots detected, as indicated above, or for each differential risk hotspot, individually. The first option is useful for comparing the level of spatial concentration that two or more types of events show along the network, or for assessing the efficiency of different hotspot detection procedures. The second option is suitable to determine which differential risk hotspot maximizes the quotient between the proportion of the type and the proportion of road length spanned, which may be a representative value of the importance of the hotspot.

\subsection{Estimating a \textit{p}-value}

Finally, assigning a statistical significance value to each differential risk hotspot is vital to reduce the possibility of focusing on certain microzones of the network that do not deserve such attention. Thus, a Monte Carlo approach was used to estimate an empirical $p$-value for each differential risk hotspot yielded by the previous step of the procedure. This approach is similar in spirit to the one proposed by \cite{bil2013identification}, which is applied in the KDE+ software mentioned before.

Getting back to previous notation, if $\{x_{1},...,x_{n}\}$ is a marked point pattern and $y_{i}$ a binary mark that indicates if $x_{i}$ corresponds, or not, to the type of event under analysis, the Monte Carlo approach implemented consists in generating $K$ simulated datasets where the locations are left fixed and the marks permutated. Concretely, this means to keep the locations $\{x_{1},...,x_{n}\}$ and to obtain a new collection of marks defined by $y_{i}^{k}=y_{\rho(i)}$, where $k$ indicates the iteration number ($k=1,...,K$) and $\rho$ is a permutation of the first $n$ natural numbers. For each simulation, the average relative probability presented by each differential risk hotspot (a weighted average per segment length of the relative probabilities estimated at the middle points of the segments composing the hotspot) is computed. At the end of the $K$ simulations, the average relative probability shown by the hotspots considering the real dataset ($\hat{r}$) is compared with the simulated ones ($\hat{s}_{k}, k=1,...,K$). The position of the real value in a numerically ordered vector formed by the simulated values and itself allows estimating an empirical $p$-value as follows:

$$p =  1 - \frac{\#\{k \in \{1,...,K\} : \hat{s}_{k} \leq \hat{r}\}}{K+1}$$

\section{Dealing with linear networks in R}

\subsection{Classes and functions}

Linear networks can be represented in R by the class \texttt{SpatialLines} of package \textsf{sp} \citep{spPackage,AppliedSpatialDataAnalysisWithR} or by simple features with package \textsf{sf} \citep{sf}. However, the class \texttt{linnet} from the \textsf{spatstat} package is the optimal for doing spatial analysis and modelling on linear networks. 

There are several functions in R that facilitate the conversion between \texttt{SpatialLines} and \texttt{linnet} objects. Specifically, \texttt{as.linnet.SpatialLines} from the \textsf{maptools} \citep{maptoolsManual} R package converts \texttt{SpatialLines} into \texttt{linnet} objects, whereas the double application (in this order) of the \texttt{as.psp} and \texttt{as.SpatialLines.psp} functions of the \textsf{spatstat} and \textsf{maptools} packages, respectively, enable the conversion from a \texttt{linnet} object into a \texttt{SpatialLines} one. 

Other useful functions available in \textsf{spatstat} for the use of linear networks are, for example, \texttt{connected.linnet} (computes the connected components of a linear network), \texttt{diameter.linnet} (computes the diameter and bounding radius of a linear network), \texttt{insertVertices} (inserts new vertices in a linear network) and \texttt{thinNetwork} (removes vertices or segments from a linear network).

\subsection{Preprocessing the linear network}

Despite not being strictly necessary, it can be convenient to preprocess the linear network in order to facilitate the subsequent statistical analysis. The \textsf{SpNetPrep} package \citep{briz2019spnetprep} can be used for this purpose. The manual edition of the network, the addition of directionality and the curation of a point pattern lying on a network can be performed through a Shiny-based application implemented in this package if the linear network represents a road network (which is the most common scenario and the one we assume for the example shown in this paper). Furthermore, \textsf{SpNetPrep} contains a function, \texttt{SimplifyLinearNetwork}, that allows users to reduce network's complexity by merging some pairs of edges of the network that fulfil certain conditions with regard to length and angle between them. In this regard, another option is to use the \texttt{gSimplify} function of \textsf{rgeos}, which provides an implementation of the Douglas-Peuker algorithm for curve simplification \citep{douglas1973algorithms}. 

\subsection{Creating a point pattern on a linear network}

Function \texttt{lpp} of package \texttt{spatstat} can be used to create an R object that represents a point pattern lying on a linear network. The \texttt{lpp} function only requires the coordinates of the events and a \texttt{linnet} object corresponding to a linear network. For instance, the following commands can be typed to create a point pattern of 100 points over the \texttt{simplenet} network provided by \textsf{spatstat}, which lies within the $[0,1]\times[0,1]$ window:

\begin{verbatim}
> x <- runif(100, 0, 1)
> y <- runif(100, 0, 1)
> simplenet.lpp <- lpp(data.frame(x, y), simplenet)
\end{verbatim}

Marks can be attached to the points forming the pattern by introducing several more columns next to the \texttt{x} and \texttt{y} coordinates. For example, one can introduce a continuous random mark following a standard normal distribution, or a categorical random mark.

\begin{verbatim}
> random_cont_mark <- rnorm(100, 0, 1)
> random_cat_mark <- letters[round(runif(100, 0, 5))+1]
> simplenet.lpp <- lpp(data.frame(x, y, random_cont_mark, random_cat_mark), 
                                  simplenet)
\end{verbatim}

In order to fit the objective of computing a relative probability for one type of event, categorical marks are required. However, recoding a continuous mark into several categories to facilitate the estimation of certain relative risk is one possible alternative.

\section{Using DRHotNet}

This section shows the complete use of the \textsf{DRHotNet} package with a dataset of crime events recorded in Chicago (Illinois, US). First of all, the following R libraries have to be loaded to reproduce the example:

\begin{verbatim}
> library(DRHotNet)
> library(lubridate)
> library(maptools)
> library(raster)
> library(rgeos)
> library(sp)
> library(SpNetPrep)
> library(spatstat)
> library(tigris)
\end{verbatim}

\subsection{Downloading and preparing the linear network}

The examples provided in this section fully employs open data available for Chicago. First, geographic data from Chicago was downloaded from the United States Census Bureau through package \textsf{tigris} \citep{walker2016tigris}. Specifically, census tracts and the road network of the state of Massachusetts were loaded into the R console. The function \texttt{intersect} from the package \textsf{raster} \citep{raster} can be used.

\begin{figure}[ht] 
\centering
\includegraphics[width=0.75\linewidth]{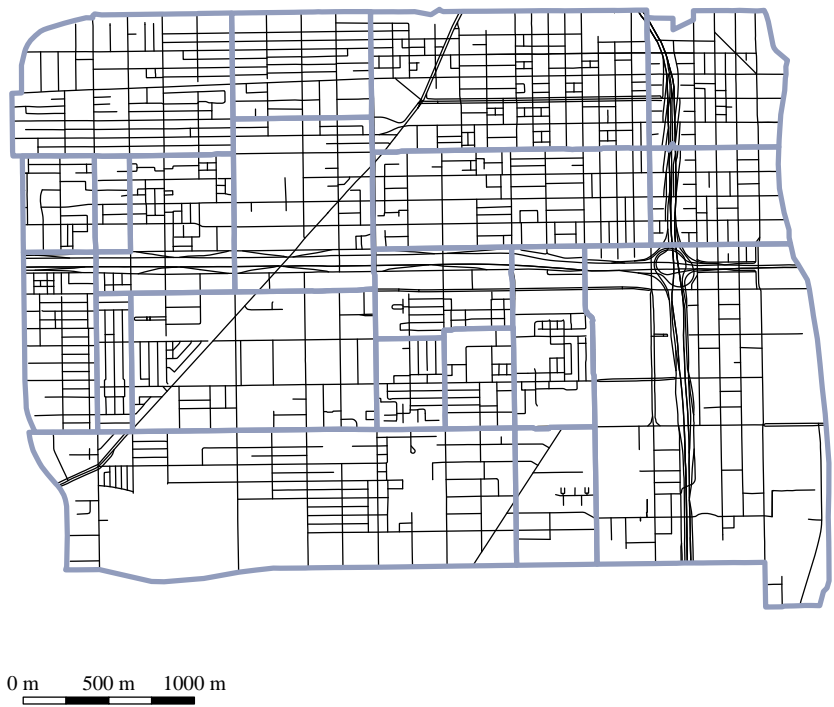} 
\caption{Road network (in black) corresponding to the Near West Side Community Area of Chicago. Census tracts of the area are overlayed in blue.}
\label{fig:ChicagoNetwork} 
\end{figure}

\begin{verbatim}
> cook.tracts <- tracts(state = "Illinois", county = "031")
> class(cook.tracts)
[1] "SpatialPolygonsDataFrame"
attr(,"package")
[1] "sp"
> cook.network <- roads(state = "Illinois", county = "031")
> class(cook.network)
[1] "SpatialLinesDataFrame"
attr(,"package")
[1] "sp"
\end{verbatim}

The objects \texttt{cook.tracts} and \texttt{cook.network} are composed of 1319 polygons and 77698 lines, respectively (as of the end of September 2019). Now, both objects are used to construct a smaller road network that corresponds to the Near West Side Community Area of Chicago.

\begin{verbatim}
> names.tracts <- as.character(cook.tracts@data[,"NAME"])
> select.tracts <- c("8378","2804","8330","2801","2808","2809","8380",
                     "8381","8331","2819","2827","2828","8382","8329",
                     "2831","2832","8333","8419","8429","2838")
> cook.tracts.select <- cook.tracts[which(names.tracts%in%select.tracts),]
> chicago.SpLines <- intersect(cook.network, cook.tracts.select)
> length(chicago.SpLines)
[1] 1116
\end{verbatim}

Object \texttt{chicago.SpLine} (\texttt{SpatialLinesDataFrame}) has 1116 lines. Then, this object's coordinates are converted into UTM (Chicago's UTM zone is 16):

\begin{verbatim}
chicago.SpLines <- spTransform(chicago.SpLines, 
                               "+proj=utm +zone=16 ellps=WGS84")
\end{verbatim}

Now the corresponding \texttt{linnet} object is created:

\begin{verbatim}
> chicago.linnet <- as.linnet(chicago.SpLines)
> chicago.linnet
Linear network with 9431 vertices and 10559 lines
Enclosing window: 
    rectangle = [442563.5, 447320] x [4634170, 4637660] units
\end{verbatim}

It is worth noting how the transformation of the network into a \texttt{linnet} object increases dramatically the number of line segments (from 1116 to 10559). This is a consequence of the fact that \texttt{SpatialLines} objects can handle curvilinear segments, made of multiple line segments, as a single line. However, \texttt{linnet} objects follow strictly the definition of linear network provided in the Introduction, which excludes this possibility.

It is required that the network is fully connected in order to allow the computation of a distance between any pair of points. This can be checked with the function \texttt{connected}.

\begin{verbatim}
> table(connected(chicago.linnet, what = "labels"))

   1    2 
9429    2
\end{verbatim}

This output means that there is a connected component of 9429 vertices and a separate component of only two vertices. The use of \texttt{connected} with the option \texttt{what = "components"} enables us to extract the larger connected component for the analysis, discarding the other one.

\begin{verbatim}
> chicago.linnet.components <- connected(chicago.linnet, 
                                         what = "components")
> chicago.linnet.components
[[1]]
Linear network with 9429 vertices and 10558 lines
Enclosing window: 
rectangle = [442563.5, 447320] x [4634170, 4637660] units
[[2]]
Linear network with 2 vertices and 1 line
Enclosing window: 
rectangle = [442563.5, 447320] x [4634170, 4637660] units
> chicago.linnet <- chicago.linnet.components[[1]]
\end{verbatim}

At this point, it is worth considering the possibility of reducing network's complexity. The function \texttt{SimplifyLinearNetwork} of \textsf{SpNetPrep} can be used for this purpose. A reasonable choice of the parameters is \texttt{Angle = 20} and \texttt{Length = 50} \citep{briz2019spnetprep}. This choice of the parameters means that a pair of segments meeting at a second-degree vertex is merged into one single segment if the angle they form (measured from 0$^{\circ}$ to 90$^{\circ}$) is lower than 20$^{\circ}$ and if the length of each of them is lower than 50 m. Hence, network's complexity is reduced (in terms of number of segments and lines) while its geometry is preserved. The following lines of code execute \texttt{SimplifyLinearNetwork} and redefine \texttt{chicago.SpLine} according to the structure of the final \texttt{linnet} object.

\begin{verbatim}
> chicago.linnet <- SimplifyLinearNetwork(chicago.linnet, 
                                          Angle = 20, Length = 50)
> chicago.linnet
Linear network with 2564 vertices and 3693 lines
Enclosing window: 
rectangle = [442563.5, 447320] x [4634170, 4637660] units
> chicago.SpLines <- as.SpatialLines.psp(as.psp(chicago.linnet))
\end{verbatim}

\begin{figure}[ht] 
  \centering
  \subfloat[]{\includegraphics[width=5cm]{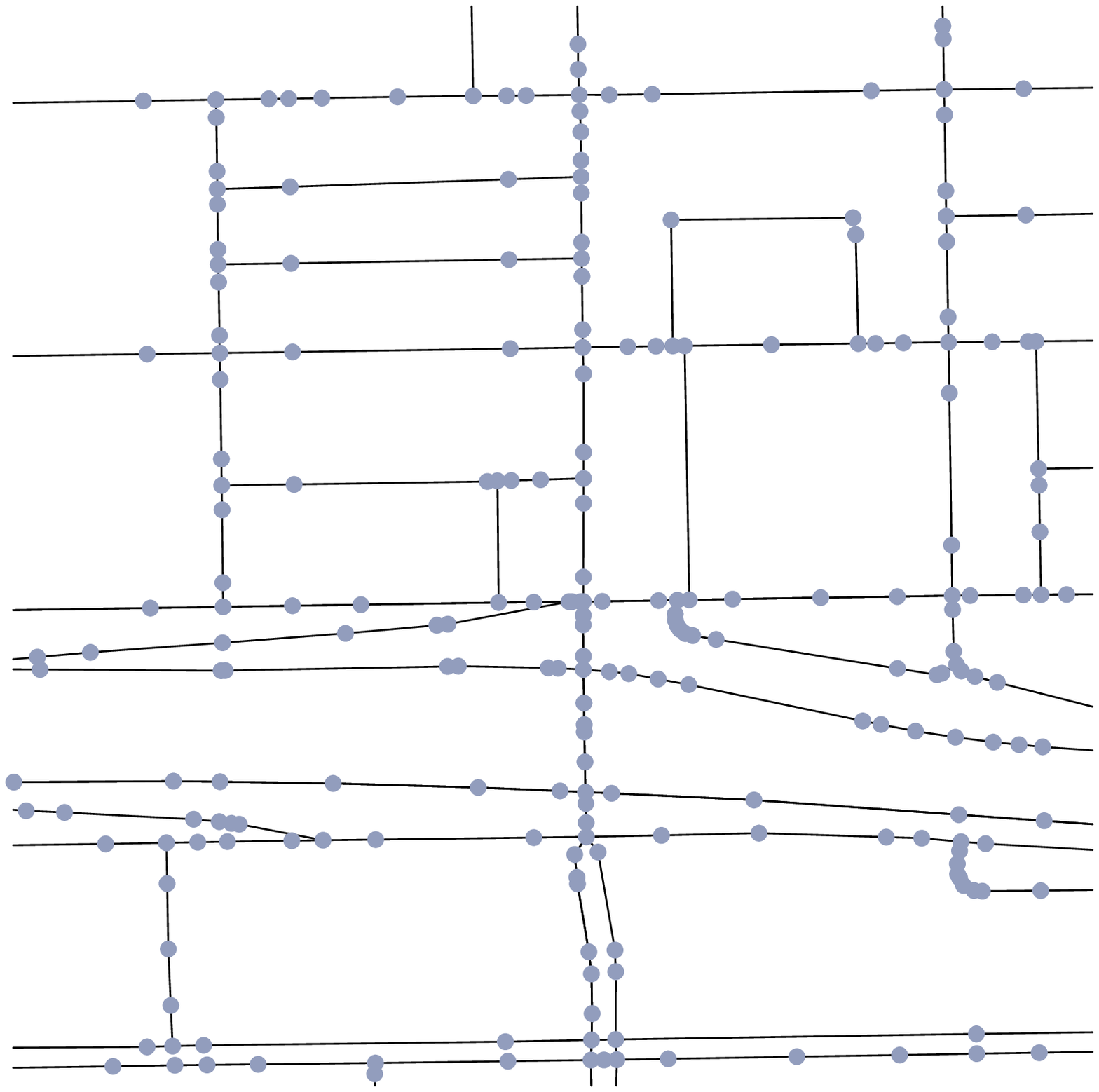}\label{fig:ChicagoNetworkZoom_a}}\hfil
  \subfloat[]{\includegraphics[width=5cm]{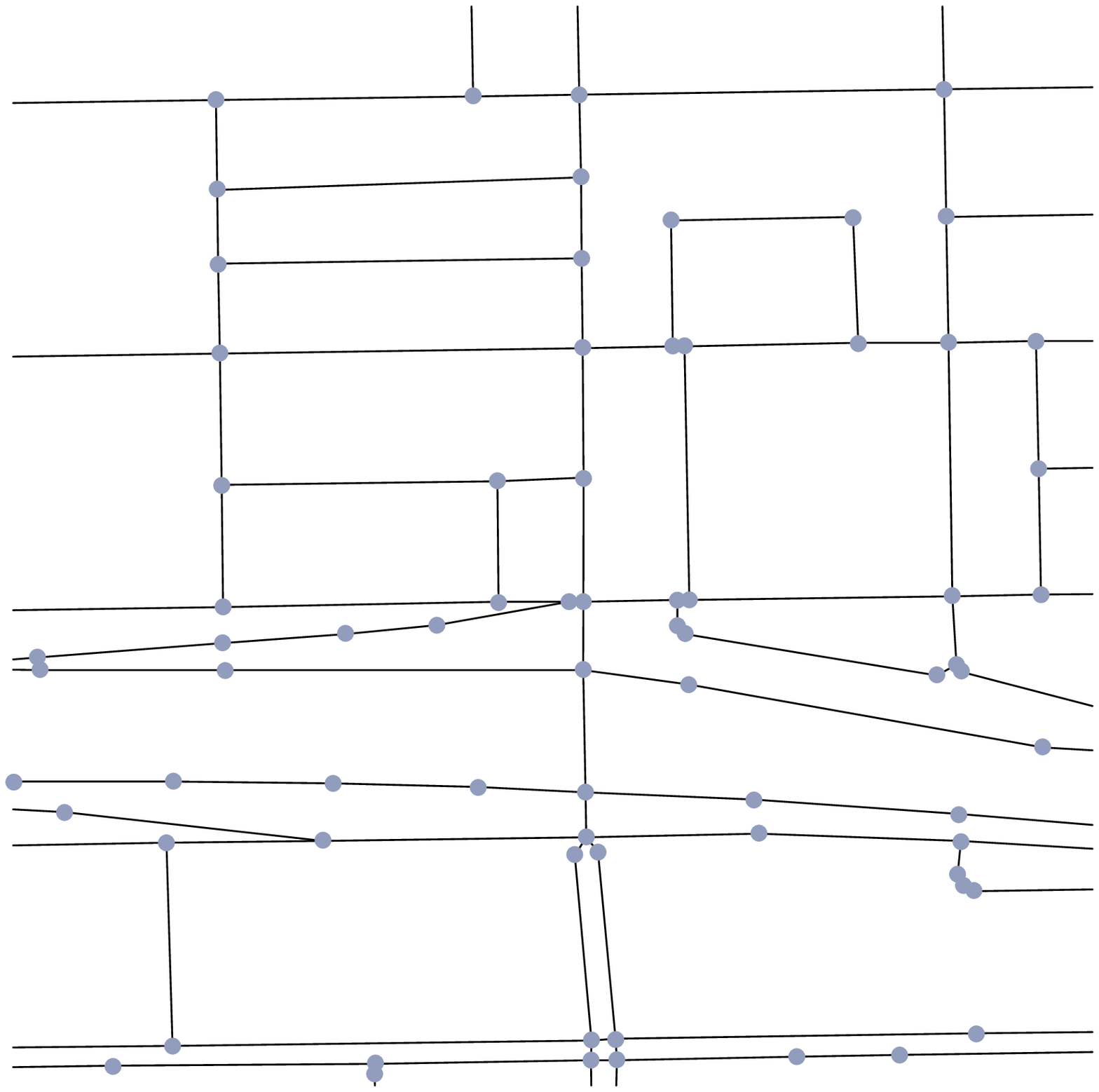}\label{fig:ChicagoNetworkZoom_b}}
  \caption{Extracting a part of the road network structure analyzed from Chicago. Original structure extracted (left), made of 273 lines and 260 vertices, and simplified version of it (right) with 114 lines and 101 vertices.}
  \label{fig:ChicagoNetworkZoom} 
\end{figure}

Thus, the final road network from Chicago that we use for the analysis has 3693 lines and 2564 vertices. An example of how \texttt{SimplifyLinearNetwork} reduces network's complexity is shown in Figure \ref{fig:ChicagoNetworkZoom}, which corresponds to a squared zone of Chicago's network with a diameter of 600 m.

\subsection{Downloading and preparing crime data}

Point-referenced crime datasets corresponding to several cities from the United States of America can be downloaded through the R package \textsf{crimedata} \citep{ashby2019studying}. Concretely, \textsf{crimedata} currently provides (as of September 2019) crime open data recorded in Austin, Boston, Chicago, Detroit, Fort Worth, Kansas City, Los Angeles, Louisville, Mesa, New York, San Francisco, Tucson and Virginia Beach. Therefore, the function \texttt{get\_crime\_data} from this package can be used for downloading a dataset of crime events recorded in Chicago in the period 2007-2018.

\begin{verbatim}
> chicago.crimes <- get_crime_data(years = 2007:2018, cities = "Chicago")
> dim(chicago.crimes)
[1] 39151    12
\end{verbatim}

The year, month and hour of occurrence of each crime can be extracted with the corresponding functions of the package \textsf{lubridate} \citep{lubridate}.

\begin{verbatim}
> chicago.crimes$year <- year(chicago.crimes$date_single)
> chicago.crimes$month <- month(chicago.crimes$date_single)
> chicago.crimes$hour <- hour(chicago.crimes$date_single)
\end{verbatim}

Then, a marked point pattern lying on \texttt{chicago.linnet} can be created with function \texttt{lpp} to provide the framework required by the \textsf{DRHotNet} package. A \texttt{data.frame} is passed to \texttt{lpp} including the coordinates of the events (in UTM), the type of event according to the receiver of the offense (\texttt{offense\_against}), and the year, month and hour of occurrence that have been just computed.

\begin{verbatim}
> chicago.crimes.coord <- data.frame(x = chicago.crimes$longitude, 
                                     y = chicago.crimes$latitude)
> coordinates(chicago.crimes.coord) <-~ x + y
> lonlat_proj <- "+proj=longlat +datum=WGS84 +ellps=WGS84 +towgs84=0,0,0"
> utm_proj <- "+proj=utm +zone=16 ellps=WGS84"
> proj4string(chicago.crimes.coord) <- lonlat_proj
> chicago.crimes.coord <- spTransform(chicago.crimes.coord, utm_proj)
> X.chicago <- lpp(data.frame(x = chicago.crimes.coord@coords[,1], 
                   y = chicago.crimes.coord@coords[,2], 
                   offense_against = chicago.crimes$offense_against,
                   year = chicago.crimes$year,
                   month = chicago.crimes$month,
                   hour = chicago.crimes$hour),
                   chicago.linnet)
Warning message:
37825 points were rejected as lying outside the specified window 
\end{verbatim}

A total of 37825 points are rejected because they lie outside the road network. Hence, a marked point pattern of 1326 crimes that lie on \texttt{chicago.linnet} remains for the analysis. The four marks are categorical, presenting the following values and absolute frequencies:

\begin{verbatim}
> table(X.chicago$data$offense_against)

   other  persons property  society 
      80      268      831      147 
> table(X.chicago$data$year)

2007 2008 2009 2010 2011 2012 2013 2014 2015 2016 2017 2018 
 138  130  133  129  116   97   94   90   85  111  105   98 
> table(X.chicago$data$month)

  1   2   3   4   5   6   7   8   9  10  11  12 
103  78 120 102 111 114 138 101 115 124 114 106 
> table(X.chicago$data$hour)

 0  1  2  3  4  5  6  7  8  9 10 11 12 13 14 15 16 17 
62 33 44 24 23 15 31 29 61 72 53 73 82 63 66 78 69 75
18  19  20  21  22  23 
52  76  70  72  52  51 
\end{verbatim}

Hence, \textsf{DRHotNet} functionalities are now applicable to \texttt{X.chicago}. In the main example shown, the relative occurrence of crimes against persons along the road network included in \texttt{X.chicago} is analyzed. To this end, the functions of the package are used following the steps that have been established for the differential risk hotspot detection methodology previously described. 

\subsection{Estimating a relative probability surface}

The function \texttt{RelativeProbabilityNetwork} of \textsf{DRHotNet} has to be used at first to estimate a relative probability surface over the linear network. As it has been explained, this implies estimating a relative probability in the middle point of each segment of the network. 

\begin{verbatim}
> rel_probs_persons <- RelativeProbabilityNetwork(X = X.chicago, 
                                                lixel_length = 50,
                                                sigma = 250,
                                                mark = "offense_against",
                                                category_mark = "persons") 
\end{verbatim}

In this example, an upper bound of 50 m is chosen for lixel's length. This means that segments shorter than 50 m are not split, whereas those longer than 50 m are split into several of no more than 50 m length. This operation is performed internally by \texttt{RelativeProbabilityNetwork} with the function \texttt{lixellate} from \textsf{spatstat}. The bandwidth parameter, $\sigma$, is set to 250 m. The \texttt{mark} and \texttt{category\_mark} parameters are used to specify the type of event that is under analysis.

\begin{figure}[ht] 
  \centering
  \subfloat[]{\includegraphics[width=7.5cm]{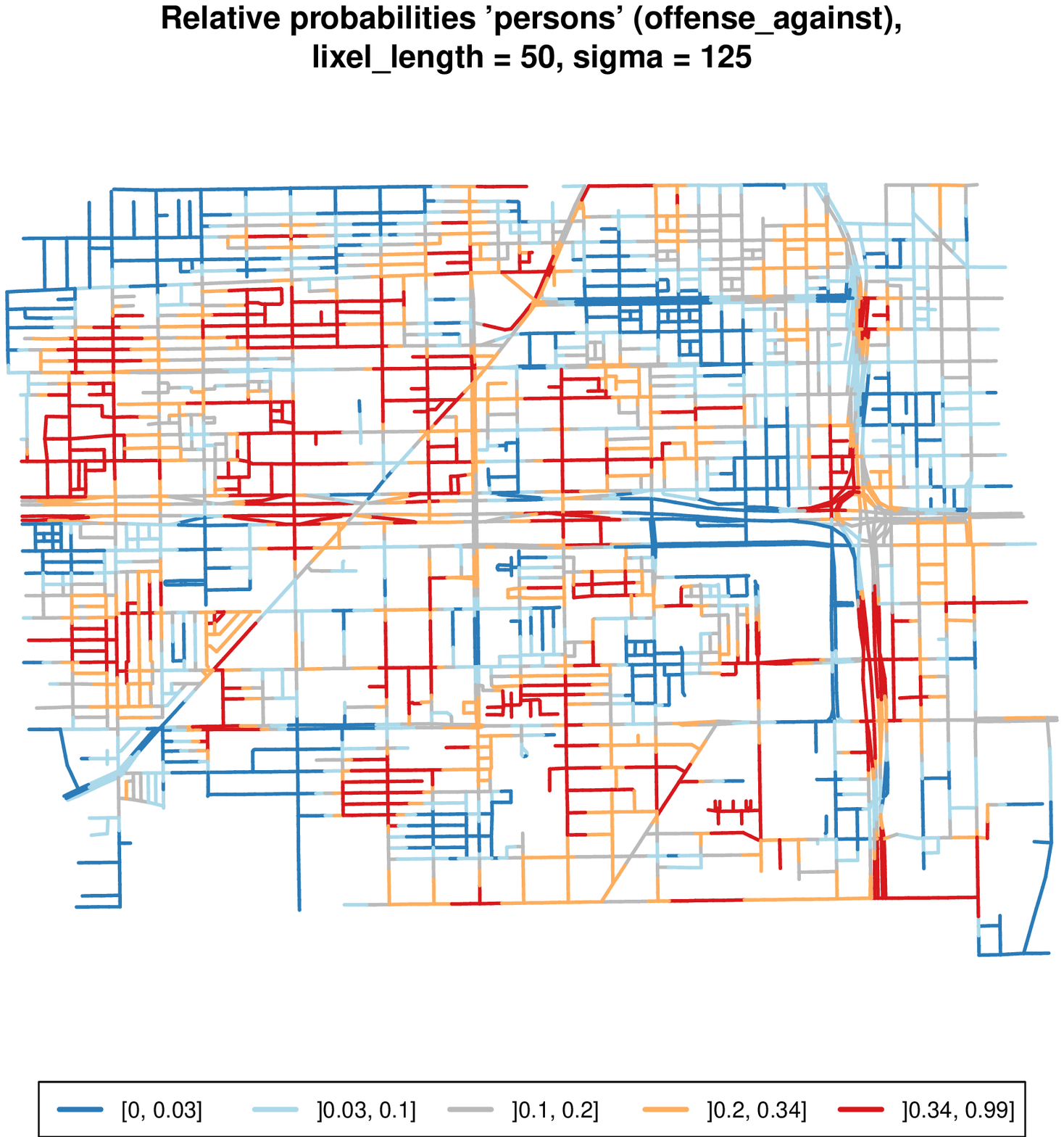}\label{fig:ChicagoNetworkRelProbs_a}}
  \subfloat[]{\includegraphics[width=7.5cm]{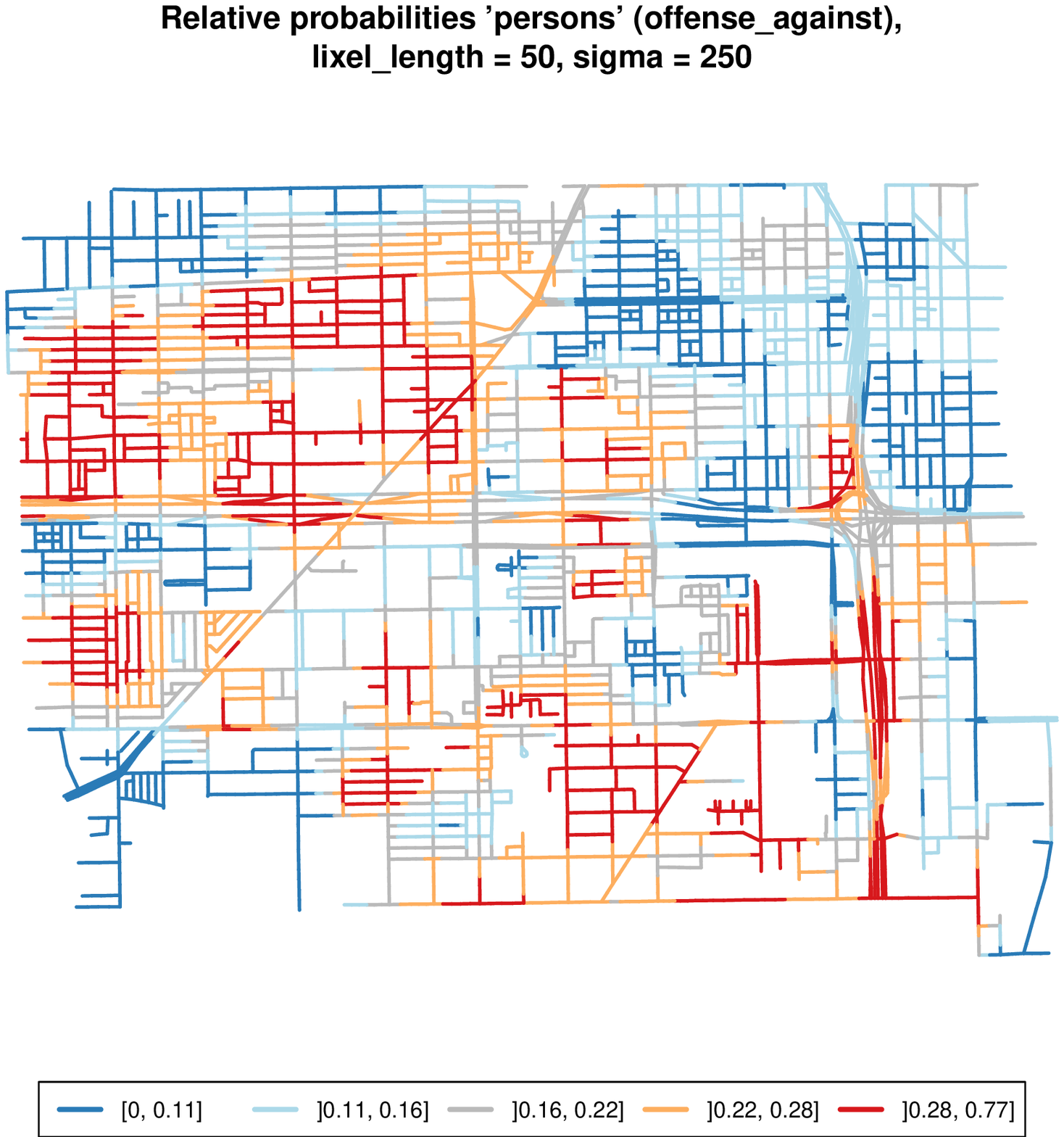}\label{fig:ChicagoNetworkRelProbs_b}}\hfil
   \subfloat[]{\includegraphics[width=7.5cm]{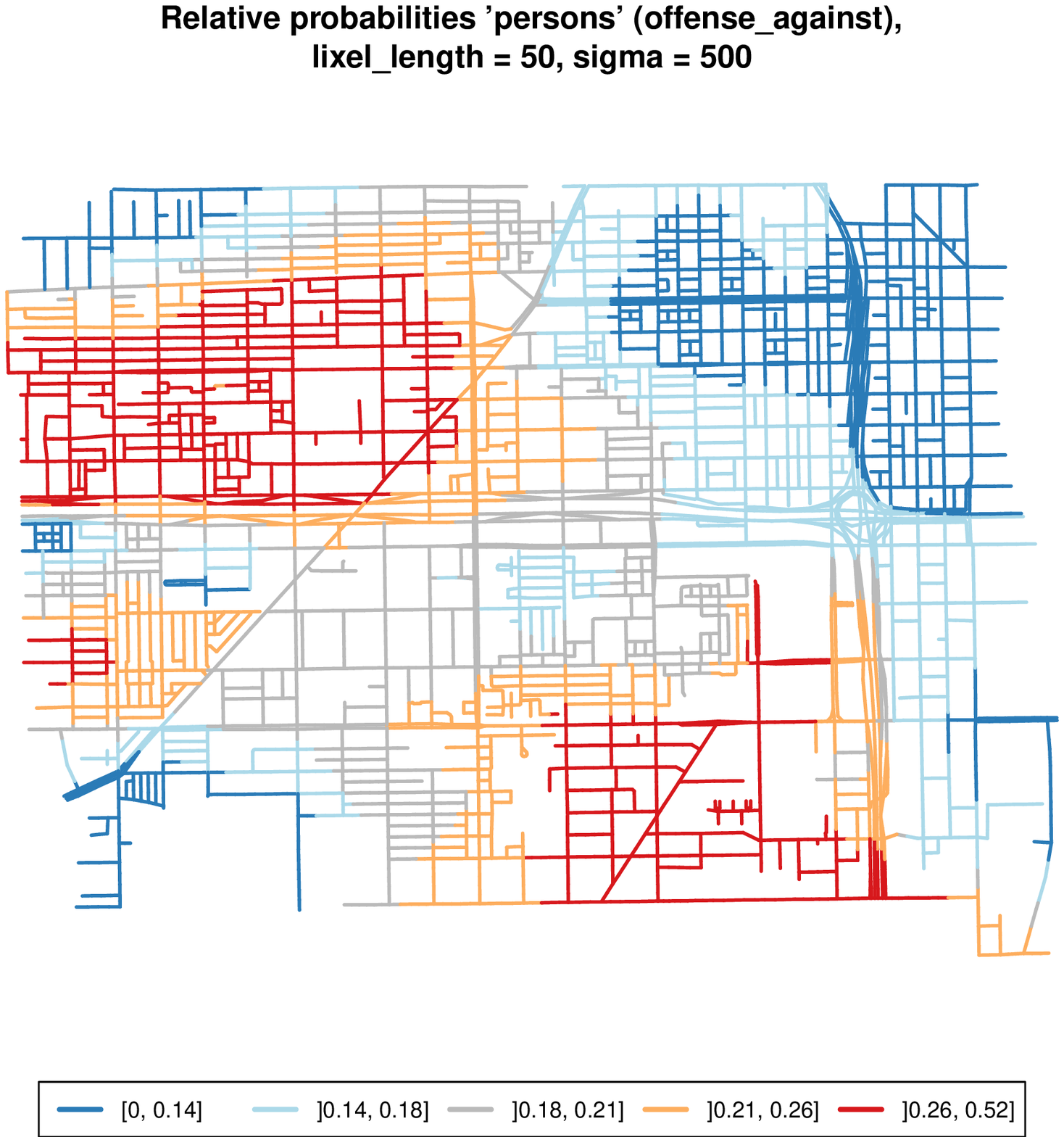}\label{fig:ChicagoNetworkRelProbs_c}}
  \subfloat[]{\includegraphics[width=7.5cm]{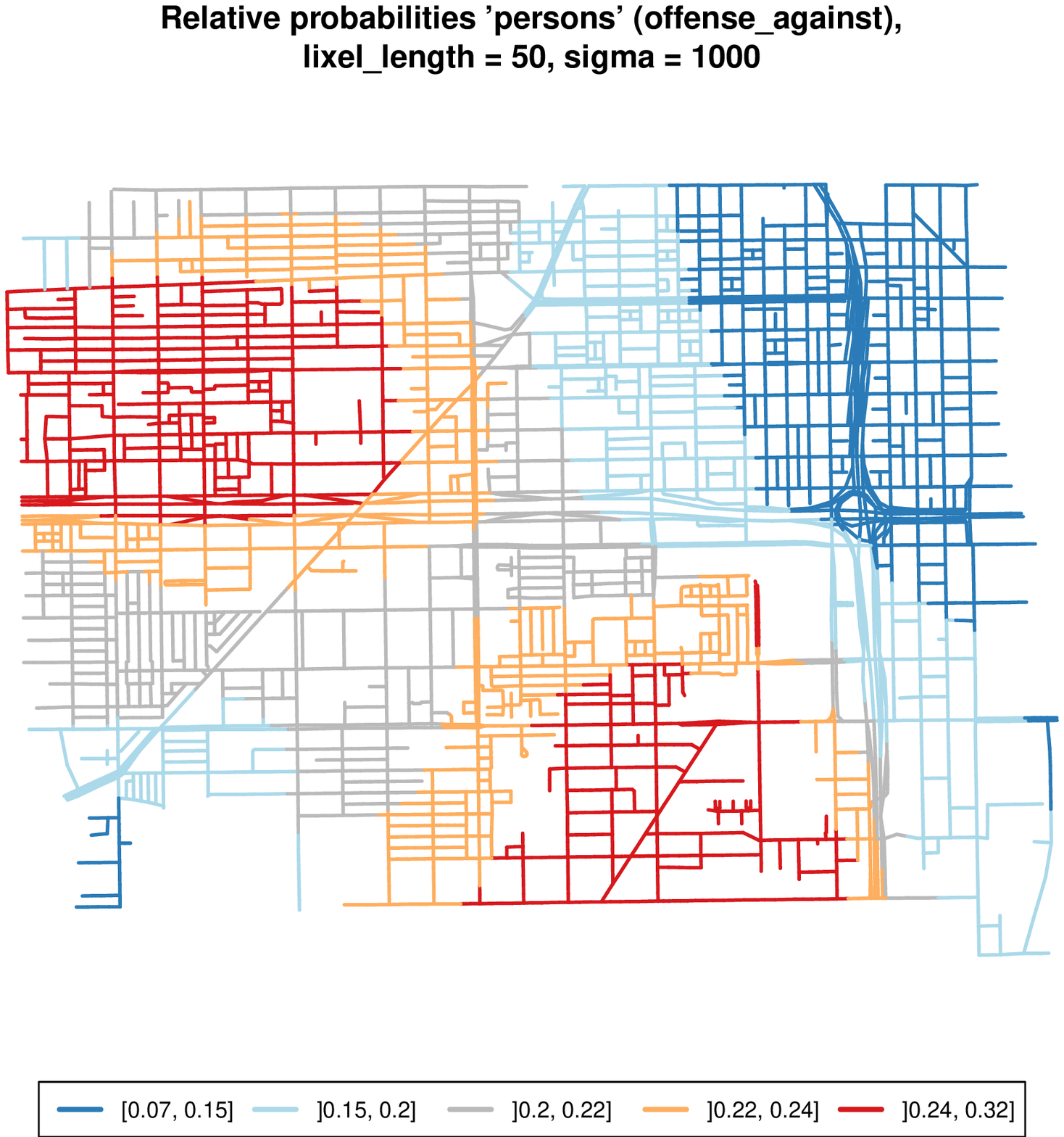}\label{fig:ChicagoNetworkRelProbs_d}}
  \caption{Outputs from the function \texttt{PlotRelativeProbabilities} for the following choices of $\sigma$: 125 (a), 250 (b), 500 (c) and 1000 (d).}
  \label{fig:ChicagoNetworkRelProbs} 
\end{figure}

The exploration of the object \texttt{rel\_probs\_persons} with function \texttt{str} allows the user to check that the choice of a 50 m threshold for lixel's length produces 7614 segments along the network. Thus, a relative probability is estimated in the middle point of each of these segments, which can be accessed typing \texttt{\$probs}:

\begin{verbatim}
> str(rel_probs_persons)
List of 5
 $ probs        : num [1:7614] 0.131 0.162 0.137 0.115 0.108 ...
 $ lixel_length : num 50
 $ sigma        : num 250
 $ mark         : chr "offense_against"
 $ category_mark: chr "persons"
\end{verbatim}

The function \texttt{PlotRelativeProbabilities} can then be used to obtain a map of the relative probability surface as the one shown in Figure \ref{fig:ChicagoNetworkRelProbs_b}. Figure \ref{fig:ChicagoNetworkRelProbs} also contains the relative probability surfaces corresponding to the choices of $\sigma=125$ (Figure \ref{fig:ChicagoNetworkRelProbs_a}), $\sigma=500$ (Figure \ref{fig:ChicagoNetworkRelProbs_c}) and $\sigma=1000$ (Figure \ref{fig:ChicagoNetworkRelProbs_d}). It can be observed that the choice of a larger value for $\sigma$ smooths the relative probability surface, which in the case of $\sigma=500$ or $\sigma=1000$ leads to the configuration of a small number of clearly distinguishable zones of the network in terms of the relative probability of offenses against persons. Indeed, whereas the use of $\sigma=125$ allows the user to obtain quite extreme relative probability values at some segments of the network (the values vary from 0 to 0.99), choosing $\sigma=1000$ causes that all the relative probabilities lie in the interval $[0.07,0.32]$. 

In view of Figure $\ref{fig:ChicagoNetworkRelProbs}$, we consider that $\sigma=250$ is a reasonable choice (although some procedures for bandwidth selection could be explored for taking this decision). Therefore, this selection of the bandwidth parameter is maintained to display now how the \texttt{DRHotspots\_k\_n} function of the package performs. 

\subsection{Detecting differential risk hotspots}

The function \texttt{DRHotspots\_k\_n} needs four parameters to be specified: \texttt{X} (a point pattern on a linear network), \texttt{rel\_probs} (an object like the one obtained in the previous step), \texttt{k} and \texttt{n}. Parameters \texttt{k} and \texttt{n} control the differential risk hotspot procedure, as it has been explained before. For example, we can try with \texttt{k = 1} and \texttt{n = 30}:

\begin{verbatim}
> hotspots_persons <- DRHotspots_k_n(X = X.chicago, 
                                     rel_probs = rel_probs_persons, 
                                     k = 1, n = 30)
\end{verbatim}

The output of the function \texttt{DRHotspots\_k\_n} presents the following structure:

\begin{verbatim}
> str(hotspots_persons)
List of 8
 $ DRHotspots   :List of 5
  ..$ : num [1:12] 247 313 314 1150 1151 ...
  ..$ : num [1:12] 551 552 553 554 5080 ...
  ..$ : num [1:8] 1093 1818 2466 4575 6127 ...
  ..$ : num 4271
  ..$ : num [1:2] 5533 5535
 $ k            : num 1
 $ n            : num 30
 $ lixel_length : num 50
 $ sigma        : num 250
 $ mark         : chr "offense_against"
 $ category_mark: chr "persons"
 $ PAI_type     : num 7.82
 \end{verbatim}
 
 The first component of \texttt{hotspots\_persons} contains the differential risk hotspots that have been detected for the values of $k$ and $n$ provided to \texttt{DRHotspots\_k\_n}. In this case, five differential risk hotspots are found, which are formed by 12, 12, 8, 1 and 2 road segments, respectively. The object \texttt{hotspots\_persons} also contains the values of the parameters involved in the computation of the hotspots and a final component that includes the global $\mathrm{PAI}_{type}$ for the set, which is 7.82 in this example. 
 
 The function \texttt{SummaryDRHotspots} can be used to provide a summary of each of the differential risk hotspots determined by \texttt{DRHotspots\_k\_n}: 
 
  \begin{verbatim}
> summary_persons <- SummaryDRHotspots(X = X.chicago, 
                                       rel_prob = rel_probs_persons,
                                       hotspots = hotspots_persons)
 \end{verbatim}
 
 The output of \texttt{SummaryDRHotspots} includes a count of the number of events located within each differential risk hotspot and how many of these correspond to the category \texttt{``persons''}:  
 
 \begin{verbatim}
> summary_persons[,c("Events type (ctr)", "All events (ctr)", 
                     "Prop. (ctr)")]
  Events type (ctr) All events (ctr) Prop. (ctr)
1                 4               11        0.36
2                 5                8        0.62
3                 0                1        0.00
4                 0                0         NaN
5                 0                0         NaN
 \end{verbatim}
 
 The summary also contains the length (in meters) of each differential risk hotspot and the $\mathrm{PAI}_{type}$ that corresponds to each of them:
 
  \begin{verbatim}
> summary_persons[,c("Length in m (ctr)", "PAI_type (ctr)")]
  Length in m (ctr) PAI_type (ctr)
1            456.86           9.13
2            366.01          14.24
3            265.71           0.00
4             42.70           0.00
5             68.78           0.00
 \end{verbatim}
 
Furthermore, the output of \texttt{SummaryDRHotspots} also provides the same statistics for an extension of each of the hotspots. The reason to include this information is because if there are not many events available in the dataset (as it happens in this example), the method can determine differential risk hotspots where very few events, if any, have taken place. Indeed, in the output of \texttt{SummaryDRHotspots} shown above, there are two hotspots including zero events, one including only one, and two more containing a very reduced number of crimes. The fact of employing kernel density estimation to infer a relative probability surface makes more convenient to think of each differential risk hotspot as the union of a center or core (what \texttt{DRHotspots\_k\_n} returns, the hotspot itself) and an extension of it. Hence, by considering an extension of the differential risk hotspot one can better appreciate the zone of the network that has been accounted for in the estimation of the relative probability values corresponding to the segments of the hotspot.
 
By default, the extension computed by \texttt{SummaryDRHotspots} coincides with a neighbourhood of the segments forming each differential risk hotspot of order $o=\frac{\sigma}{\mathrm{Lixel\;length}}$ (rounded to the nearest integer), although a different order can be specified through the parameter \texttt{order\_extension}. In this example, we have $o=\frac{250}{50}=5$, which is used by \texttt{SummaryDRHotspots} if no other order is indicated: 
 
 \begin{verbatim}
> summary_persons[,c("Events type (ext)", "All events (ext)", 
                     "Prop. (ext)")]
  Events type (ext) All events (ext) Prop. (ext)
1                14               42        0.33
2                14               50        0.28
3                11               33        0.33
4                10               30        0.33
5                 5               15        0.33
> summary_persons[,c("Length in m (ext)", "PAI_type (ext)")]
  Length in m (ext) PAI_type (ext)
1           3526.69           4.14
2           3285.57           4.44
3           1921.33           5.97
4           1318.60           7.91
5           1072.17           4.86
 \end{verbatim}
 
It can be observed that all extensions of the differential risk hotspots include a reasonable number of events and that the corresponding proportions of offenses against persons are clearly above the global proportion for the dataset, which is $268/1326\approx0.20$.

\subsection{Assessing the statistical significance of the hotspots}

Following with the choice of \texttt{k = 1} and \texttt{n = 30}, it only remains to estimate a $p$-value for each of the differential risk hotspots detected. This can be done calling the function \texttt{SummaryDRHotspots} again and specifying \texttt{compute\_p\_value = T}. A total of 200 iterations are selected for performing the Monte Carlo simulation process:

\begin{verbatim}
> summary_persons <- SummaryDRHotspots(X = X.chicago, 
                                       rel_prob = rel_probs_persons,
                                       hotspots = hotspots_persons,
                                       compute_p_value = T,
                                       n_it = 200)
> summary_persons$`p-value`
[1] 0.015 0.010 0.035 0.035 0.000
\end{verbatim}

Therefore, the five differential risk hotspots detected with $k=1$ and $n=30$ are statistically significant ($p<0.05$). It is worth noting, however, that the usual significance level of 0.05 should be reduced (corrected) if many differential risk hotspots are detected to avoid the presence of multiple comparison problems.

\subsection{Choosing \textit{k} and \textit{n}}

\begin{figure}[ht] 
  \centering
  \subfloat[]{\includegraphics[width=7.5cm]{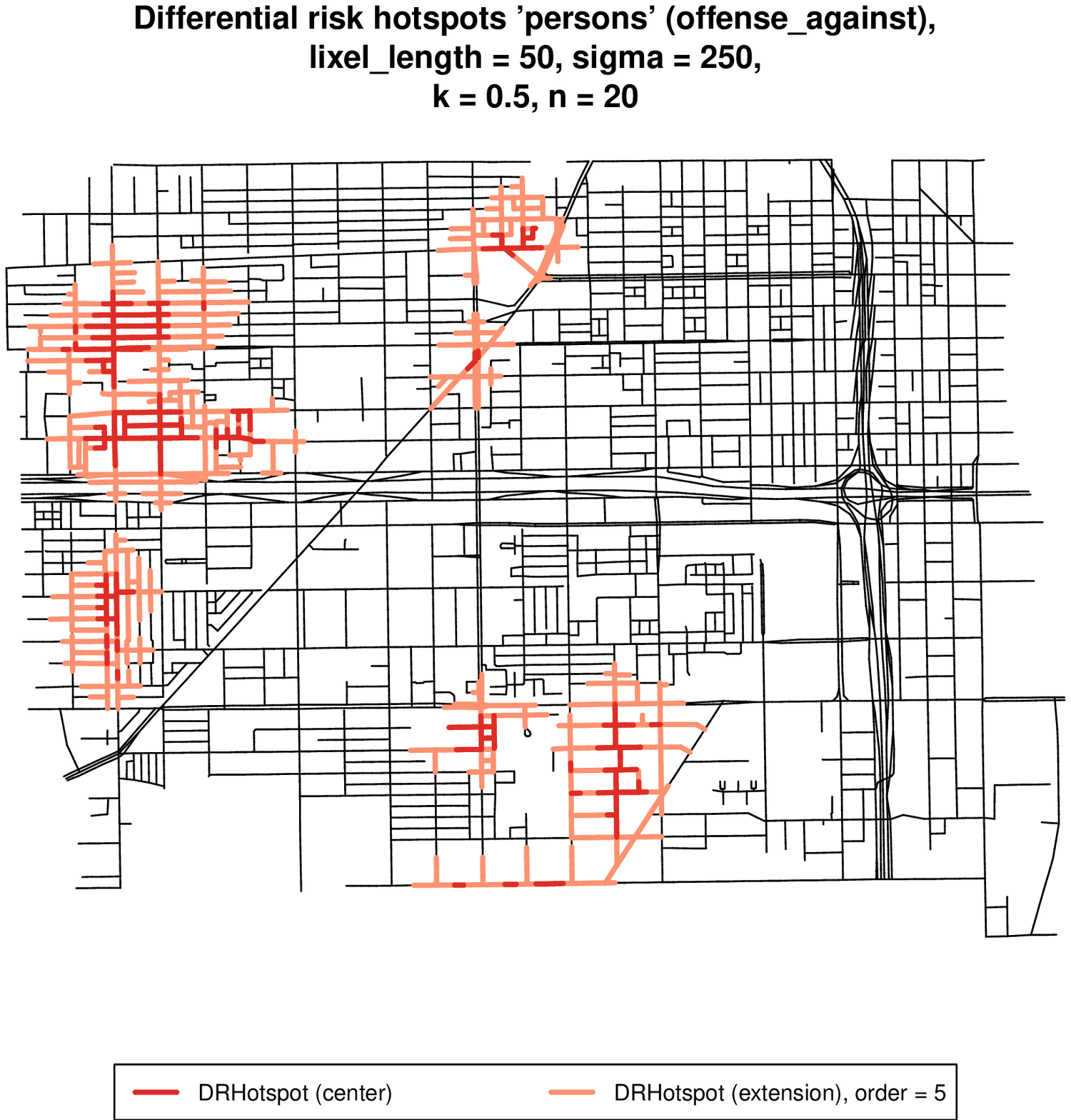}\label{fig:ChicagoNetworkHotspots_a}}
  \subfloat[]{\includegraphics[width=7.5cm]{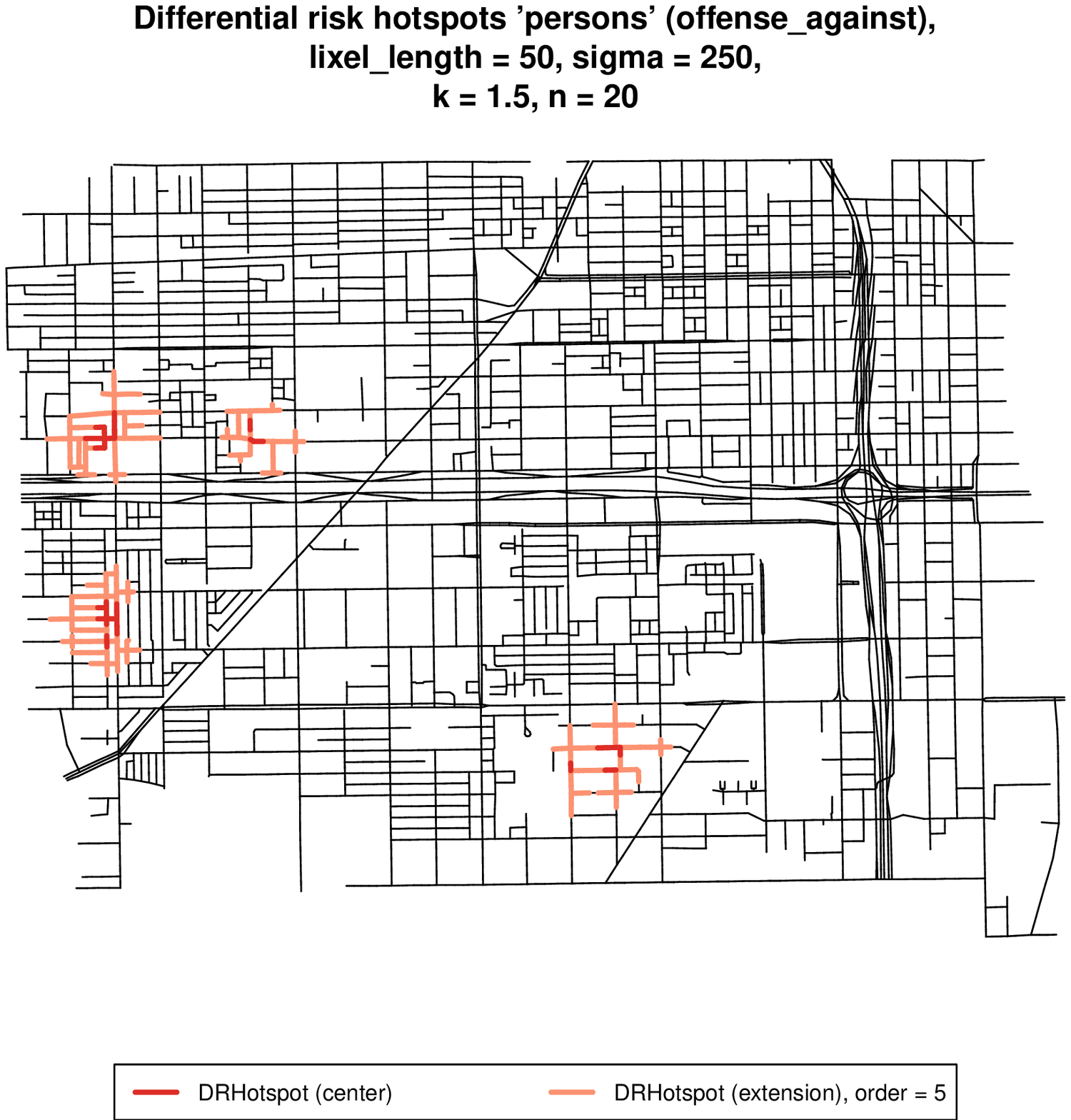}\label{fig:ChicagoNetworkHotspots_b}}\hfil
   \subfloat[]{\includegraphics[width=7.5cm]{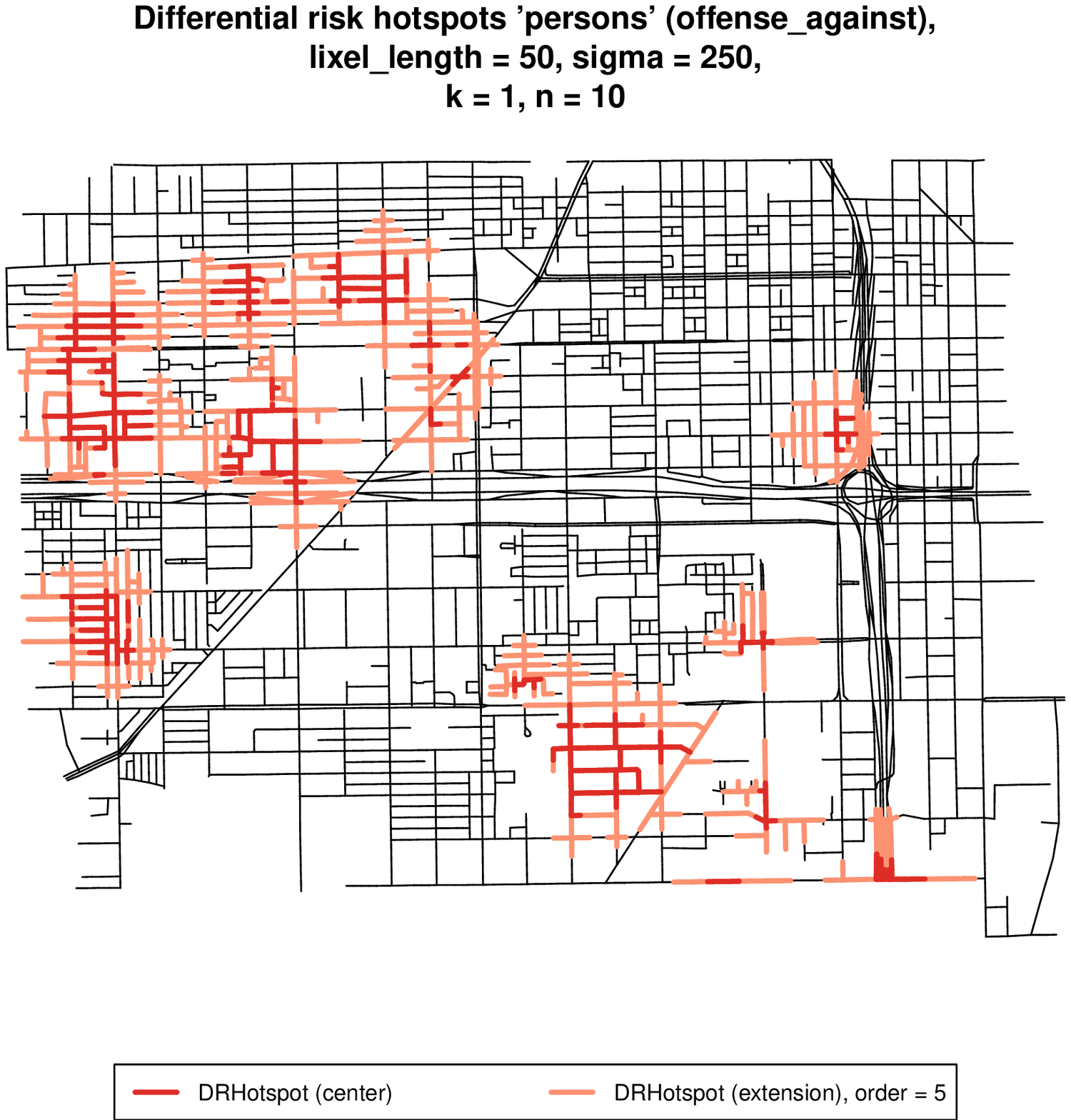}\label{fig:ChicagoNetworkHotspots_c}}
  \subfloat[]{\includegraphics[width=7.5cm]{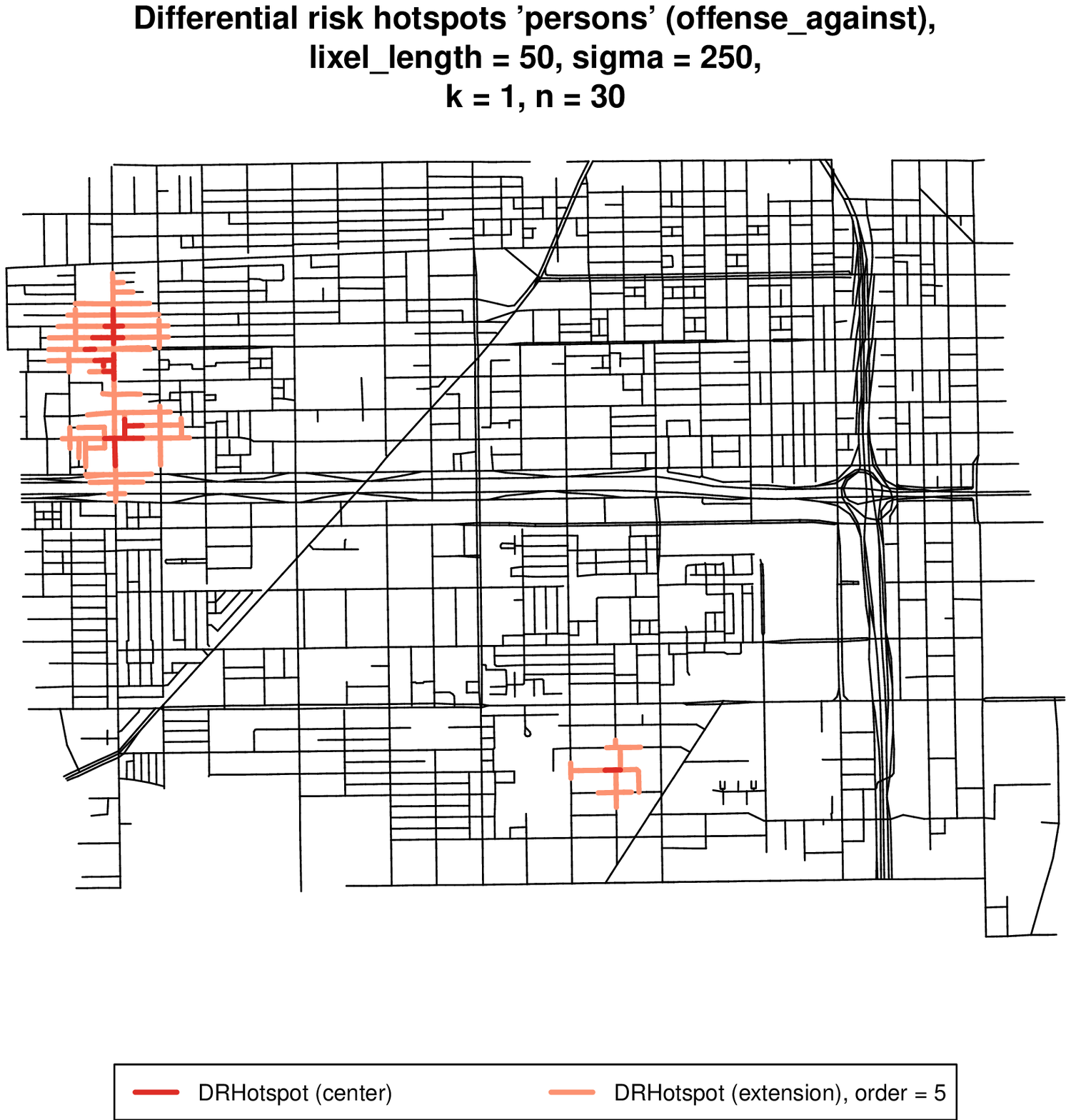}\label{fig:ChicagoNetworkHotspots_d}}
  \caption{Outputs from the function \texttt{PlotHotspots} for the following choices of $k$ and $n$: $k=0.5$ and $n=20$ (a), $k=1.5$ and $n=20$ (b), $k=1$ and $n=10$ (c) and $k=1$ and $n=30$ (d).}
  \label{fig:ChicagoNetworkHotspots} 
\end{figure}

Remember that a higher value of either \texttt{k} or \texttt{n} represents using a more stringent criterion regarding hotspot detection. This is illustrated through the four maps available in Figure \ref{fig:ChicagoNetworkHotspots}, which can be generated with the function \texttt{PlotHotspots} of \textsf{DRHotNet}. For instance, as in the following example for the object \texttt{hotspots} created previously (which corresponds to Figure \ref{fig:ChicagoNetworkHotspots_d}):

\begin{verbatim}
> PlotHotspots(X = X.chicago, hotspots = hotspots)
\end{verbatim}

Indeed, Figure \ref{fig:ChicagoNetworkHotspots} shows the differential risk hotspots that \texttt{DRHotspots\_k\_n} detects for the choices of \texttt{k = 0.5} and \texttt{n = 20} (Figure \ref{fig:ChicagoNetworkHotspots_a}), \texttt{k = 1.5} and \texttt{n = 20} (Figure \ref{fig:ChicagoNetworkHotspots_b}), \texttt{k = 1} and \texttt{n = 10} (Figure \ref{fig:ChicagoNetworkHotspots_c}), and \texttt{k = 1} and \texttt{n = 30} (Figure \ref{fig:ChicagoNetworkHotspots_d}). Two of these combinations of conditions on $k$ and $n$ are implicitly represented by the other two. Consequently, the differential risk hotspots shown in Figure \ref{fig:ChicagoNetworkHotspots_b} are contained in Figure \ref{fig:ChicagoNetworkHotspots_a}, and those in Figure \ref{fig:ChicagoNetworkHotspots_d} are contained in Figure \ref{fig:ChicagoNetworkHotspots_c}. The choice of \texttt{k = 1} and \texttt{n = 30} leads to the highest global $\mathrm{PAI}_{type}$ among the four combinations of parameters indicated, with the value of 15.6 mentioned before. In this regard, we recommend performing a sensitivity analysis on the values of $k$ and $n$ to decide which combination is more convenient. The sensitivity analysis carried out by \cite{briz2019hotspots} yielded that a choice around $k=1.5$ and $n=45$ was optimal in terms of the $\mathrm{PAI}_{type}$ for the traffic accident dataset that was investigated. However, each dataset should require a specific analysis. 

Thus, a sensitivity analysis on $k$ and $n$ can be carried out with the function \texttt{Sensitivity\_k\_n}. The user has to provide a point pattern (\texttt{X}), an object containing the relative probabilities of a type of event along the network (\texttt{rel\_probs}) and a set of values for $k$ and $n$ (\texttt{ks} and \texttt{ns}, respectively):

\begin{verbatim}
> sensitivity_analysis <- Sensitivity_k_n(X = X.chicago, 
                                          rel_prob = rel_probs_persons, 
                                          ks = seq(0,3,0.5), 
                                          ns = seq(10,30,5))
> sensitivity_analysis
        n = 10 n = 15 n = 20 n = 25 n = 30
k = 0     3.05   3.60   4.36   4.72   5.58
k = 0.5   3.93   4.90   5.96   6.23   6.52
k = 1     4.62   5.82   6.11   6.66   7.82
k = 1.5   5.44   7.63   4.52   0.00   0.00
k = 2     5.66  18.14   0.00   0.00     NA
k = 2.5   7.06     NA     NA     NA     NA
k = 3     0.00     NA     NA     NA     NA
\end{verbatim}

The output from \texttt{Sensitivity\_k\_n} is a matrix that contains the $\mathrm{PAI}_{type}$ values that correspond to each combination of $k$ and $n$ indicated by \texttt{ks} and \texttt{ns}. A \texttt{NA} value represents that no differential risk hotspots can be found for such combination of parameters. According to this matrix, the highest $\mathrm{PAI}_{type}$ value is achieved for $k=2$ and $n=15$. 

Therefore, one can choose the values of $k$ and $n$ that maximize the $\mathrm{PAI}_{type}$ (considering the parameters provided in \texttt{ks} and \texttt{ns}) to determine the final set of differential risk hotspots. However, this criteria may lead sometimes to detect a very low number of differential risks hotspots and hence to miss zones of the network that may also deserve some attention. Hence, a more conservative approach could be considering several combinations of $k$ and $n$ that yield some of the highest values of $\mathrm{PAI}_{type}$ and explore each set of differential risk hotspots associated. Then, one could investigate the output of \texttt{SummaryDRHotspots} for each combination of parameters (including the computation of $p$-values) to better decide which zones of the network are relevant for the type of event of interest.

\subsection{Other applications of the methodology}

This final section shows the results that are obtained for other type of events for comparative purposes. First, the marks \texttt{X.chicago} are recoded into binary outcomes as follows:

\begin{verbatim}
> year_after_2012 <- ifelse(X.chicago$data$year>2012, "Yes", "No")
> month_winter <- ifelse(X.chicago$data$month%in%c(12,1,2), "Yes", "No")
> hour_21_3 <- ifelse(X.chicago$data$hour%in%c(21:23,0:3), "Yes", "No")
> marks(X.chicago) <- data.frame(as.data.frame(marks(X.chicago)),
                               year_after_2012 = year_after_2012,
                               month_winter = month_winter,
                               hour_21_3 = hour_21_3)
\end{verbatim}

The relative probability surfaces have to be computed. We used the same values for \texttt{lixel.length} and \texttt{sigma} than in the previous examples.

\begin{verbatim}
> rel_probs_after_2012 <- RelativeProbabilityNetwork(X = X.chicago, 
                                                  lixel_length = 50,
                                                  sigma = 250,
                                                  mark = "year_after_2012",
                                                  category_mark = "Yes") 
> rel_probs_winter <- RelativeProbabilityNetwork(X = X.chicago, 
                                                 lixel_length = 50,
                                                 sigma = 250,
                                                 mark = "month_winter",
                                                 category_mark = "Yes") 
> rel_probs_21_3 <- RelativeProbabilityNetwork(X = X.chicago, 
                                               lixel_length = 50,
                                               sigma = 250,
                                               mark = "hour_21_3",
                                               category_mark = "Yes") 
\end{verbatim}

The corresponding sensitivity analyses are carried out:

\begin{verbatim}
> sensitivity_after_2012 <- Sensitivity_k_n(X = X.chicago, 
                                           rel_prob = rel_probs_after_2012, 
                                           ks = seq(0,3,0.5), 
                                           ns = seq(10,30,5))
> sensitivity_after_2012
        n = 10 n = 15 n = 20 n = 25 n = 30
k = 0     2.09   2.67   3.45   4.98   8.71
k = 0.5   2.58   2.99   4.28   7.16  16.81
k = 1     5.43   7.61   9.71  17.64  28.91
k = 1.5   3.40     NA     NA     NA     NA
k = 2       NA     NA     NA     NA     NA
k = 2.5     NA     NA     NA     NA     NA
k = 3       NA     NA     NA     NA     NA
> sensitivity_winter <- Sensitivity_k_n(X = X.chicago, 
                                       rel_prob = rel_probs_winter, 
                                       ks = seq(0,3,0.5), 
                                       ns = seq(10,30,5))
> sensitivity_winter
        n = 10 n = 15 n = 20 n = 25 n = 30
k = 0     2.88   4.02   5.06   7.36   9.89
k = 0.5   3.14   4.60   4.30   6.97   6.03
k = 1     5.18   7.35   1.32   0.00     NA
k = 1.5   0.83   0.00     NA     NA     NA
k = 2     0.00     NA     NA     NA     NA
k = 2.5     NA     NA     NA     NA     NA
k = 3       NA     NA     NA     NA     NA
> sensitivity_21_3 <- Sensitivity_k_n(X = X.chicago, 
                                     rel_prob = rel_probs_21_3, 
                                     ks = seq(0,3,0.5), 
                                     ns = seq(10,30,5))
> sensitivity_21_3
        n = 10 n = 15 n = 20 n = 25 n = 30
k = 0     2.42   3.20   4.49   4.69   4.81
k = 0.5   3.00   4.08   5.30   5.55   6.03
k = 1     3.04   3.94   4.74   4.11   0.00
k = 1.5   3.67   4.58   6.37   5.94     NA
k = 2     5.46   5.79   7.30   5.58     NA
k = 2.5  12.03  11.36  14.41   5.67     NA
k = 3     0.00   0.00   0.00     NA     NA
\end{verbatim}

\begin{figure}[ht] 
  \centering
  \subfloat[]{\includegraphics[width=5cm]{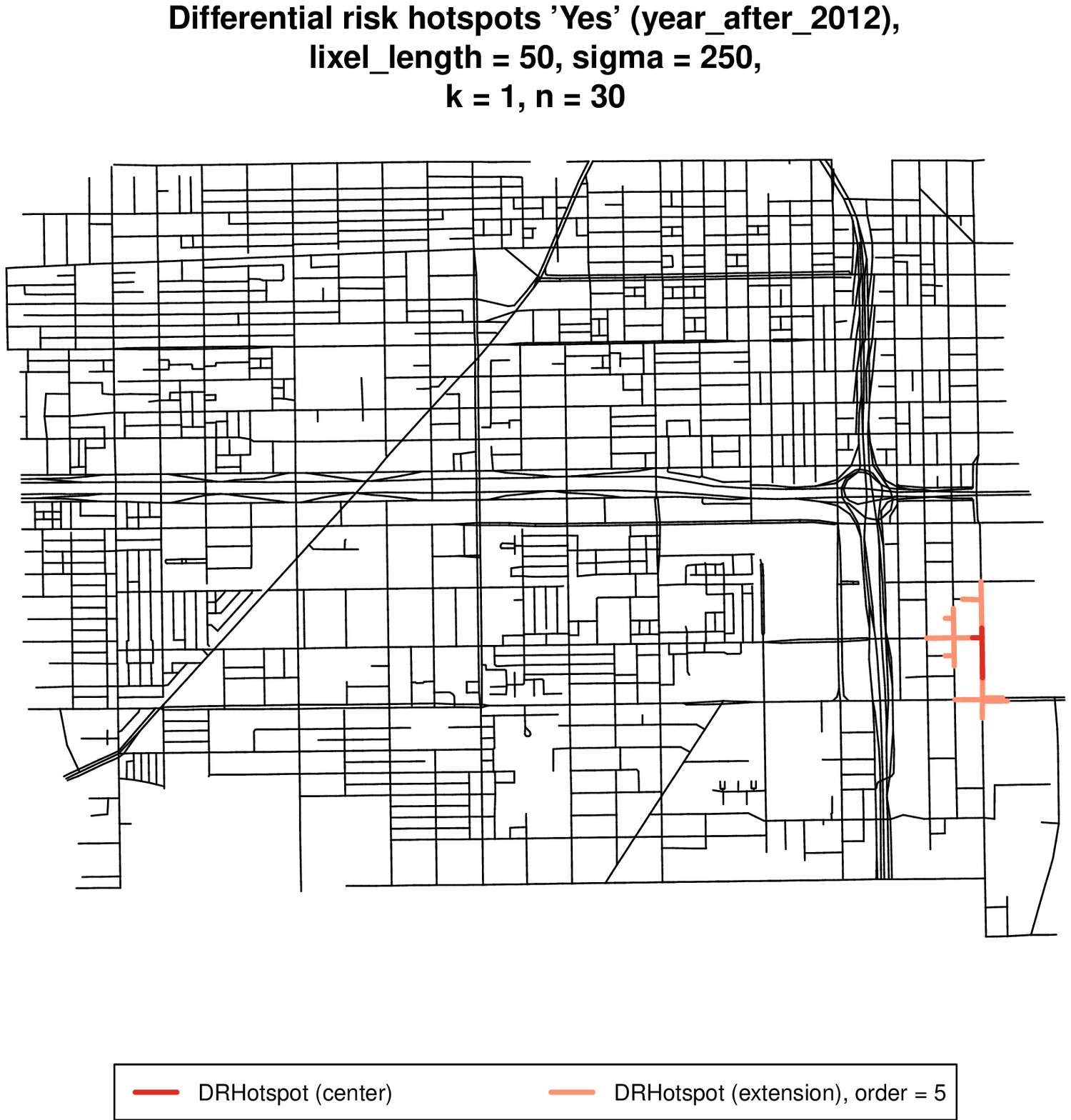}\label{fig:Comparison_a}}
  \subfloat[]{\includegraphics[width=5cm]{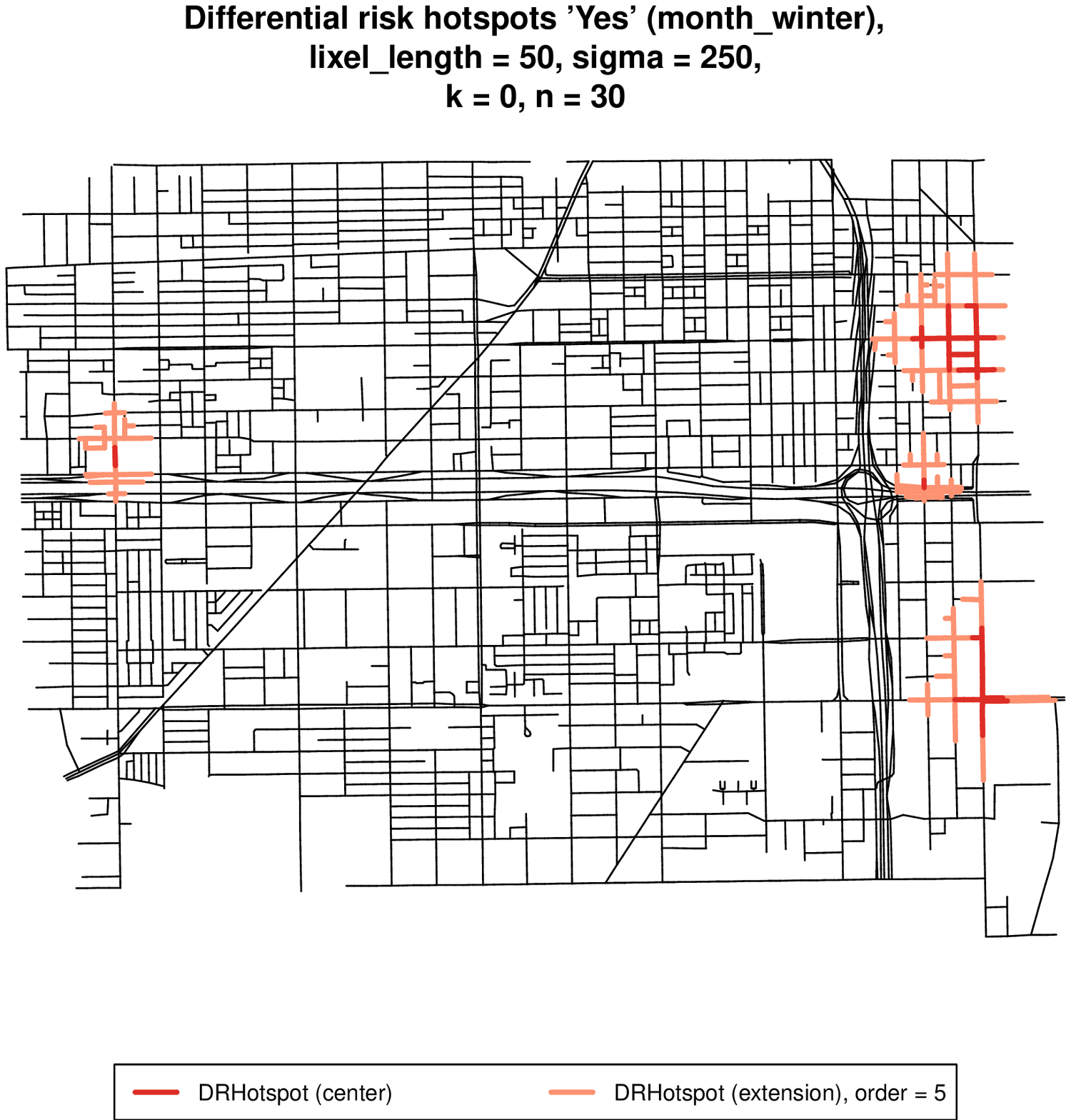}\label{fig:Comparison_b}}
   \subfloat[]{\includegraphics[width=5cm]{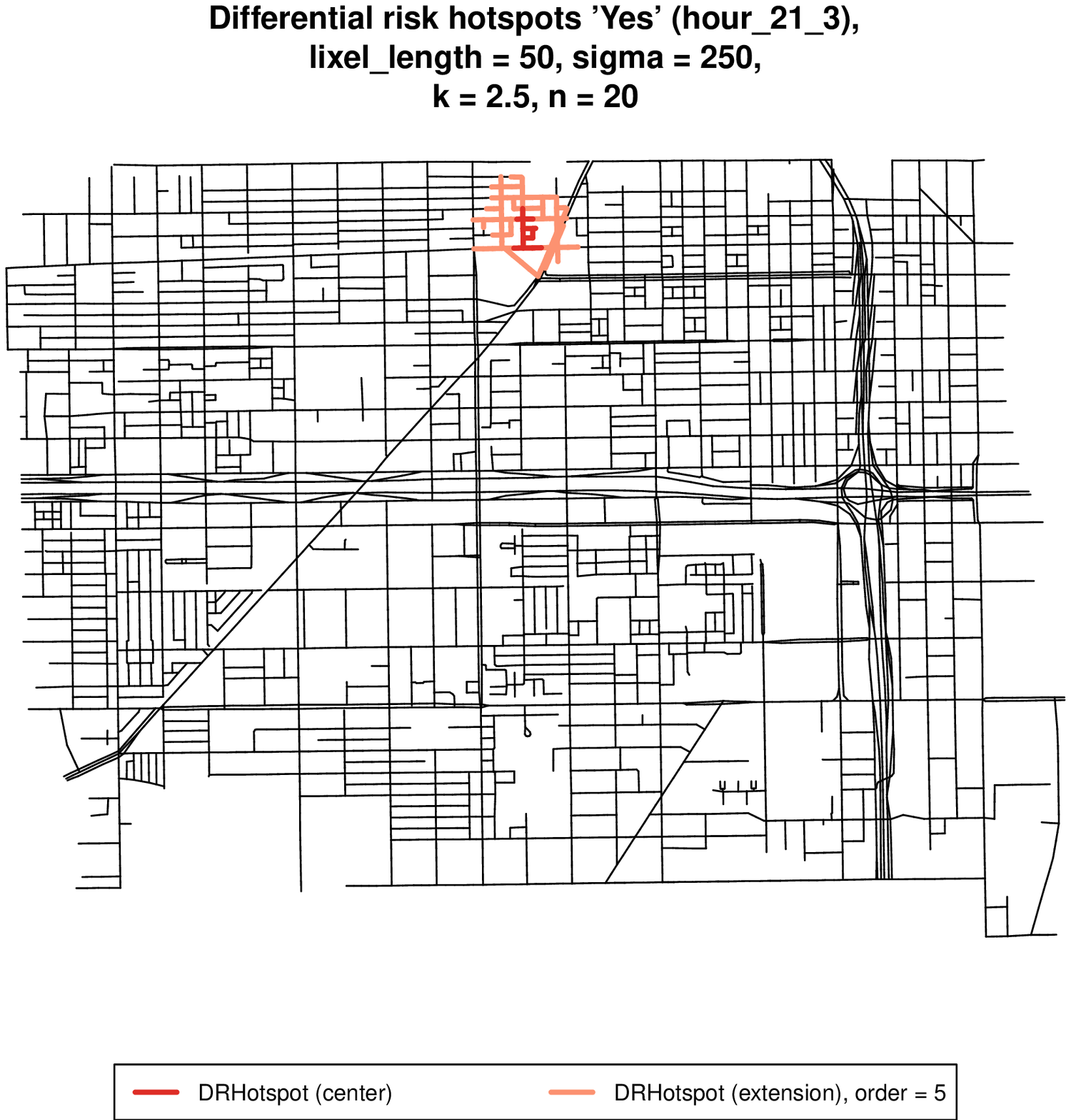}\label{fig:Comparison_c}}
  \caption{Outputs from the function \texttt{PlotHotspots} for the marks \texttt{year\_after\_2012}, \texttt{month\_winter} and \texttt{hour\_21\_3} and the categorical value \texttt{Yes} for the three, considering the combinations of $k$ and $n$ that maximize the $\mathrm{PAI}_{type}$ (for the values of $k$ and $n$ tested).}
  \label{fig:Comparison} 
\end{figure}

The highest $\mathrm{PAI}_{type}$ values for \texttt{year\_after\_2012}, \texttt{month\_winter} and \texttt{hour\_21\_3} are 28.91, 9.89 and 14.41, respectively. The differential risk hotspots that are obtained for the combination of $k$ and $n$ that lead to these $\mathrm{PAI}_{type}$ values can be visualized with \texttt{PlotHotspots} (Figure \ref{fig:Comparison}):

\begin{verbatim}
> hotspots_after_2012 <- DRHotspots_k_n(X = X.chicago, 
                                        rel_prob = rel_probs_after_2012, 
                                        k = 1, 
                                        n = 30)
> PlotHotspots(X = X.chicago, hotspots_after_2012)

> hotspots_winter <- DRHotspots_k_n(X = X.chicago, 
                                    rel_prob = rel_probs_winter, 
                                    k = 0, 
                                    n = 30)
> PlotHotspots(X = X.chicago, hotspots_winter)

> hotspots_21_3 <- DRHotspots_k_n(X = X.chicago, 
                                  rel_prob = rel_probs_21_3, 
                                  k = 2.5, 
                                  n = 20)
> PlotHotspots(X = X.chicago, hotspots_21_3)
\end{verbatim}

\section{Summary}

The R package \textsf{DRHotNet} for detecting differential risk hotspots on linear networks has been described. The use of linear networks in the context of hotspot detection is becoming more important over the years, particularly in the fields of criminology and traffic safety. In addition, it is also of great interest sometimes to detect zones of a linear network where certain type of event is especially overrepresented. Hence, \textsf{DRHotNet} consists of an easy-to-use tool implemented in R to accurately locate the microzones of a linear network where the incidence of a type of event is considerably higher than in the rest of it.

\clearpage

\bibliography{bibliography}

\begin{thebibliography}{}

\bibitem[All{\'e}vius, 2018]{allevius2018scanstatistics}
All{\'e}vius, B. (2018).
\newblock scanstatistics: Space-time anomaly detection using scan statistics.
\newblock {\em The Journal of Open Source Software}, 3:515.

\bibitem[Andresen et~al., 2017]{andresen2017trajectories}
Andresen, M.~A., Curman, A.~S., and Linning, S.~J. (2017).
\newblock The trajectories of crime at places: understanding the patterns of
  disaggregated crime types.
\newblock {\em Journal of Quantitative Criminology}, 33(3):427--449.

\bibitem[Ang et~al., 2012]{ang2012geometrically}
Ang, Q.~W., Baddeley, A., and Nair, G. (2012).
\newblock Geometrically corrected second order analysis of events on a linear
  network, with applications to ecology and criminology.
\newblock {\em Scandinavian Journal of Statistics}, 39(4):591--617.

\bibitem[Anselin, 1995]{anselin1995local}
Anselin, L. (1995).
\newblock Local indicators of spatial association—{LISA}.
\newblock {\em Geographical Analysis}, 27(2):93--115.

\bibitem[Ashby, 2019]{ashby2019studying}
Ashby, M.~P. (2019).
\newblock {Studying Crime and Place with the Crime Open Database: Social and
  Behavioural Scienes}.
\newblock {\em Research Data Journal for the Humanities and Social Sciences},
  1(aop):1--16.

\bibitem[Baddeley et~al., 2017]{baddeley2017stationary}
Baddeley, A., Nair, G., Rakshit, S., and McSwiggan, G. (2017).
\newblock “{S}tationary” point processes are uncommon on linear networks.
\newblock {\em Stat}, 6(1):68--78.

\bibitem[Baddeley et~al., 2015]{baddeley2015spatial}
Baddeley, A., Rubak, E., and Turner, R. (2015).
\newblock {\em Spatial point patterns: methodology and applications with R}.
\newblock CRC Press.

\bibitem[B{\'\i}l et~al., 2013]{bil2013identification}
B{\'\i}l, M., Andr{\'a}{\v{s}}ik, R., and Jano{\v{s}}ka, Z. (2013).
\newblock Identification of hazardous road locations of traffic accidents by
  means of kernel density estimation and cluster significance evaluation.
\newblock {\em Accident Analysis \& Prevention}, 55:265--273.

\bibitem[B{\'\i}l et~al., 2016]{bil2016kde+}
B{\'\i}l, M., Andr{\'a}{\v{s}}ik, R., Svoboda, T., and Sedon{\'\i}k, J. (2016).
\newblock {The KDE+ software: a tool for effective identification and ranking
  of animal-vehicle collision hotspots along networks}.
\newblock {\em Landscape ecology}, 31(2):231--237.

\bibitem[Bivand and Lewin-Koh, 2017]{maptoolsManual}
Bivand, R. and Lewin-Koh, N. (2017).
\newblock {\em maptools: {T}ools for {R}eading and {H}andling {S}patial
  {O}bjects}.
\newblock R package version 0.9-2.

\bibitem[Bivand et~al., 2013]{AppliedSpatialDataAnalysisWithR}
Bivand, R.~S., Pebesma, E., and Gómez-Rubio, V. (2013).
\newblock {\em Applied spatial data analysis with {R}, Second edition}.
\newblock Springer, NY.

\bibitem[Briz-Red{\'o}n, 2019]{briz2019spnetprep}
Briz-Red{\'o}n, {\'A}. (2019).
\newblock {SpNetPrep: An R package using Shiny to facilitate spatial statistics
  on road networks}.
\newblock {\em Research Ideas and Outcomes}, 5:e33521.

\bibitem[Briz-Red{\'o}n et~al., 2019a]{briz2019hotspots}
Briz-Red{\'o}n, {\'A}., Mart{\'\i}nez-Ruiz, F., and Montes, F. (2019a).
\newblock Identification of differential risk hotspots for collision and
  vehicle type in a directed linear network.
\newblock {\em Accident Analysis \& Prevention}, 132:105278.

\bibitem[Briz-Red{\'o}n et~al., 2019b]{briz2019spatial}
Briz-Red{\'o}n, {\'A}., Mart{\'\i}nez-Ruiz, F., and Montes, F. (2019b).
\newblock Spatial analysis of traffic accidents near and between road
  intersections in a directed linear network.
\newblock {\em Accident Analysis \& Prevention}, 132:105252.

\bibitem[Chainey et~al., 2008]{chainey2008utility}
Chainey, S., Tompson, L., and Uhlig, S. (2008).
\newblock The utility of hotspot mapping for predicting spatial patterns of
  crime.
\newblock {\em Security Journal}, 21(1-2):4--28.

\bibitem[Deng et~al., 2019]{deng2019density}
Deng, M., Yang, X., Shi, Y., Gong, J., Liu, Y., and Liu, H. (2019).
\newblock A density-based approach for detecting network-constrained clusters
  in spatial point events.
\newblock {\em International Journal of Geographical Information Science},
  33(3):466--488.

\bibitem[Douglas and Peucker, 1973]{douglas1973algorithms}
Douglas, D.~H. and Peucker, T.~K. (1973).
\newblock Algorithms for the reduction of the number of points required to
  represent a digitized line or its caricature.
\newblock {\em Cartographica: the International Journal for Geographic
  Information and Geovisualization}, 10(2):112--122.

\bibitem[Drawve and Wooditch, 2019]{drawve2019research}
Drawve, G. and Wooditch, A. (2019).
\newblock A research note on the methodological and theoretical considerations
  for assessing crime forecasting accuracy with the predictive accuracy index.
\newblock {\em Journal of Criminal Justice}, page 101625.

\bibitem[Eckardt and Mateu, 2017]{eckardt2017second}
Eckardt, M. and Mateu, J. (2017).
\newblock Second-order and local characteristics of network intensity
  functions.
\newblock {\em arXiv preprint arXiv:1712.01555}.

\bibitem[Eckardt and Mateu, 2018]{eckardt2018point}
Eckardt, M. and Mateu, J. (2018).
\newblock Point patterns occurring on complex structures in space and
  space-time: An alternative network approach.
\newblock {\em Journal of Computational and Graphical Statistics},
  27(2):312--322.

\bibitem[Getis and Ord, 1992]{getis1992analysis}
Getis, A. and Ord, J. (1992).
\newblock {The Analysis of Spatial Association by Use of Distance Statistics}.
\newblock {\em Geographical Analysis}, 24(3).

\bibitem[Grolemund and Wickham, 2011]{lubridate}
Grolemund, G. and Wickham, H. (2011).
\newblock {Dates and Times Made Easy with {lubridate}}.
\newblock {\em Journal of Statistical Software}, 40(3):1--25.

\bibitem[Gómez-Rubio et~al., 2005]{gomez2005detecting}
Gómez-Rubio, V., Ferrándiz-Ferragud, J., and López-Quílez, A. (2005).
\newblock Detecting clusters of disease with {R}.
\newblock {\em Journal of Geographical Systems}, 7(2):189--206.

\bibitem[Gómez-Rubio et~al., 2019]{JSSv090i14}
Gómez-Rubio, V., Moraga, P., Molitor, J., and Rowlingson, B. (2019).
\newblock Dclusterm: Model-based detection of disease clusters.
\newblock {\em Journal of Statistical Software, Articles}, 90(14):1--26.

\bibitem[Hijmans, 2019]{raster}
Hijmans, R.~J. (2019).
\newblock {\em raster: Geographic Data Analysis and Modeling}.
\newblock R package version 2.8-19.

\bibitem[Kelsall and Diggle, 1995a]{kelsall1995kernel}
Kelsall, J.~E. and Diggle, P.~J. (1995a).
\newblock Kernel estimation of relative risk.
\newblock {\em Bernoulli}, 1(1-2):3--16.

\bibitem[Kelsall and Diggle, 1995b]{kelsall1995non}
Kelsall, J.~E. and Diggle, P.~J. (1995b).
\newblock Non-parametric estimation of spatial variation in relative risk.
\newblock {\em Statistics in medicine}, 14(21-22):2335--2342.

\bibitem[Kelsall and Diggle, 1998]{kelsall1998spatial}
Kelsall, J.~E. and Diggle, P.~J. (1998).
\newblock Spatial variation in risk of disease: a nonparametric binary
  regression approach.
\newblock {\em Journal of the Royal Statistical Society: Series C (Applied
  Statistics)}, 47(4):559--573.

\bibitem[Kulldorff, 1997]{kulldorff1997spatial}
Kulldorff, M. (1997).
\newblock A spatial scan statistic.
\newblock {\em Communications in Statistics-Theory and methods},
  26(6):1481--1496.

\bibitem[Lloyd, 2010]{lloyd2010spatial}
Lloyd, C. (2010).
\newblock {\em Spatial data analysis: an introduction for GIS users}.
\newblock Oxford University Press.

\bibitem[McSwiggan et~al., 2017]{mcswiggan2017kernel}
McSwiggan, G., Baddeley, A., and Nair, G. (2017).
\newblock Kernel density estimation on a linear network.
\newblock {\em Scandinavian Journal of Statistics}, 44(2):324--345.

\bibitem[McSwiggan et~al., 2019]{mcswiggan2019estimation}
McSwiggan, G., Baddeley, A., and Nair, G. (2019).
\newblock Estimation of relative risk for events on a linear network.
\newblock {\em Statistics and Computing}, pages 1--16.

\bibitem[Meyer et~al., 2017]{meyer2017spatio}
Meyer, S., Held, L., and H{\"o}hle, M. (2017).
\newblock Spatio-temporal analysis of epidemic phenomena using the r package
  surveillance.
\newblock {\em Journal of Statistical Software}, 77(11).

\bibitem[Moradi et~al., 2019]{moradi2019resample}
Moradi, M.~M., Cronie, O., Rubak, E., Lachieze-Rey, R., Mateu, J., and
  Baddeley, A. (2019).
\newblock {Resample-smoothing of Voronoi intensity estimators}.
\newblock {\em Statistics and Computing}, 29(5):995–1010.

\bibitem[Moradi et~al., 2018]{moradi2018kernel}
Moradi, M.~M., Rodr{\'\i}guez-Cort{\'e}s, F.~J., and Mateu, J. (2018).
\newblock On kernel-based intensity estimation of spatial point patterns on
  linear networks.
\newblock {\em Journal of Computational and Graphical Statistics},
  27(2):302--311.

\bibitem[Nie et~al., 2015]{nie2015network}
Nie, K., Wang, Z., Du, Q., Ren, F., and Tian, Q. (2015).
\newblock A network-constrained integrated method for detecting spatial cluster
  and risk location of traffic crash: {A} case study from {W}uhan, {C}hina.
\newblock {\em Sustainability}, 7(3):2662--2677.

\bibitem[Pebesma, 2018]{sf}
Pebesma, E. (2018).
\newblock {Simple Features for R: Standardized Support for Spatial Vector
  Data}.
\newblock {\em {The R Journal}}.

\bibitem[Pebesma and Bivand, 2005]{spPackage}
Pebesma, E.~J. and Bivand, R.~S. (2005).
\newblock Classes and methods for spatial data in {R}.
\newblock {\em R News}, 5(2):9--13.

\bibitem[Rakshit et~al., 2019a]{rakshit2019efficient}
Rakshit, S., Baddeley, A., and Nair, G. (2019a).
\newblock {Efficient Code for Second Order Analysis of Events on a Linear
  Network}.
\newblock {\em Journal of Statistical Software}, 90(1):1--37.

\bibitem[Rakshit et~al., 2019b]{rakshitfast}
Rakshit, S., Davies, T., Moradi, M.~M., McSwiggan, G., Nair, G., Mateu, J., and
  Baddeley, A. (2019b).
\newblock {Fast Kernel Smoothing of Point Patterns on a Large Network using
  Two-dimensional Convolution}.
\newblock {\em International Statistical Review}.

\bibitem[Schnell et~al., 2017]{schnell2017influence}
Schnell, C., Braga, A.~A., and Piza, E.~L. (2017).
\newblock {The influence of community areas, neighborhood clusters, and street
  segments on the spatial variability of violent crime in Chicago}.
\newblock {\em Journal of Quantitative Criminology}, 33(3):469--496.

\bibitem[Steenbeek and Weisburd, 2016]{steenbeek2016action}
Steenbeek, W. and Weisburd, D. (2016).
\newblock {Where the action is in crime? An examination of variability of crime
  across different spatial units in The Hague, 2001--2009}.
\newblock {\em Journal of Quantitative Criminology}, 32(3):449--469.

\bibitem[Walker, 2016]{walker2016tigris}
Walker, K. (2016).
\newblock {tigris: An R package to access and work with geographic data from
  the US Census Bureau}.
\newblock {\em The R Journal}, 8(2):231--242.

\bibitem[Weisburd, 2015]{weisburd2015law}
Weisburd, D. (2015).
\newblock The law of crime concentration and the criminology of place.
\newblock {\em Criminology}, 53(2):133--157.

\bibitem[Xie and Yan, 2008]{xie2008kernel}
Xie, Z. and Yan, J. (2008).
\newblock Kernel density estimation of traffic accidents in a network space.
\newblock {\em Computers, Environment and Urban systems}, 32(5):396--406.

\bibitem[Xie and Yan, 2013]{xie2013detecting}
Xie, Z. and Yan, J. (2013).
\newblock Detecting traffic accident clusters with network kernel density
  estimation and local spatial statistics: an integrated approach.
\newblock {\em Journal of Transport Geography}, 31:64--71.

\end{thebibliography}

\end{document}